\documentclass[letter,oneside,9pt,twocolumn,article]{aiaa}
\usepackage{graphicx}
\usepackage{rotating}
\usepackage{bm}
\usepackage{flushend}
\usepackage{cite}
\usepackage[bf,footnotesize]{subfigure}
\usepackage{adjustbox}
\usepackage{tabularx}
\usepackage{multirow}
\usepackage{multicol}
\usepackage{indentfirst} 
\usepackage[hidelinks]{hyperref}
\usepackage{float}
\usepackage{xcolor}
\usepackage[round]{natbib}
\setcitestyle{authoryear,open={(},close={)}} 

\usepackage{mwe}
\usepackage[markcase=noupper]{scrlayer-scrpage}
\usepackage{caption} 

\ihead*{{\it Physics of Fluids 37, 096141 (2025), \href{https://doi.org/10.1063/5.0285170}{10.1063/5.0285170}}}
\ohead*{\pagemark}
\ifoot*{\rm }
\ofoot*{\rm Unrestricted content}
\newcommand{\alb}{\vspace{0.1cm}\\} 
\newcommand{\mfd}{\displaystyle}
\renewcommand{\vec}[1]{\bm{#1}}

\renewcommand{\fontsizetable}{\footnotesize\scalefont{1.0}}

\renewcommand{\vec}[1]{\bm{#1}}
\author{Felipe Martin Rodriguez Fuentes\thanks{Graduate Student}~~and~
Bernard Parent\thanks{Associate Professor, bparent@arizona.edu.}\\[0.3em] \it University of Arizona, Tucson, AZ 85721, USA.
}

\title{
Vibrational-Electron Heating in Plasma Flows:\\ A Thermodynamically Consistent Model
}

\abstract{ 
Accurate prediction of electron temperature ($T_{\rm e}$) in non-equilibrium plasma flows is critical, yet hampered by inadequate models for electron heating from vibrationally excited states. Prior models often relied on ad-hoc scaling or flawed applications of detailed balance that failed to ensure the convergence of electron temperature and species-specific vibrational temperature ($T_{\rm v}$) at thermal equilibrium. This paper introduces a novel, thermodynamically consistent electron heating model derived rigorously from the principle of detailed balance. By assuming a Boltzmann vibrational distribution and employing an effective activation energy, our approach yields a simple heating-to-cooling ratio of $\exp(\theta_{\rm v}/T_{\rm e}-\theta_{\rm v}/T_{\rm v})$, where $\theta_{\rm v}$ is the characteristic vibrational temperature of {the species under consideration}. This formulation guarantees that $T_{\rm e}$ correctly converges to $T_{\rm v}$ at equilibrium. A key advantage is that our model can utilize total cooling rates determined from swarm experiments, leading to higher accuracy at low electron temperatures. For re-entry flows, the proposed approach predicts an electron temperature several times lower than previous models which results in improved agreement with some flight test data.  These more reliable predictions can significantly enhance the modeling fidelity of plasma-assisted combustion, laser-induced plasmas, and various hypersonic plasma technologies such as electron transpiration cooling or magnetohydrodynamic force generators.
}

\nomenclature{
\begin{nomenclaturelist}{Roman symbols}

\item[$C_k$]{charge of $k$th  species, C}
\item[$C_{\rm i}$]{charge of ion species, C}
\item[$C_{\rm e}$]{charge of electrons, C}
\item[$E_{\rm i}$]{electric field along $i$th dimension, V/m}
\item[$\vec{E}$]{electric field vector, V/m}
\item[$E^\star$]{reduced electric field, $E/N$, $\rm Vm^2$}
\item[$E_{\rm exp}^\star$]{reduced electric field from swarm experimental data, $E/N$, $\rm Vm^2$}
\item[$E_{\rm bol}^\star$]{reduced electric field from BOLSIG+ data, $E/N$, $\rm Vm^2$}
\item[$E_k^\star$]{reduced electric field of $k$th neutral species, $\rm Vm^2$}
\item[$\mathcal{E}_{kl}$]{ activation energy of the $l$th electron impact process of the $k$th species, J/particle}
\item[$\mathcal{E}_{\rm eff}$]{ effective activation energy of vibrational excitation processes, J/particle}
\item[$\mathcal{E}_{{\rm N_2}l}$]{activation energy of nitrogen vib. excitation to $n$th level, J/particle}
\item[$\mathcal{E}_{n}$]{ activation energy of the $n$th nitrogen vibrational excitation process, J/particle}
\item[$e_{\rm e}$]{electron specific internal energy, J/kg}
\item[$e_{\rm t}$]{total specific internal energy, J/kg}
\item[$e_{\rm v}$]{nitrogen average specific vibrational energy (excluding the zero-point energy), J/kg}
\item[$e_{\rm v}^0$]{nitrogen average specific vibrational energy in equilibrium (excluding the zero-point energy), J/kg}
\item[$f_v$]{fraction of nitrogen in $v$th vibrational excited level}
\item[$f_{v=0}$]{fraction of nitrogen in ground state}
\item[$h_{k}^{0}$]{heat of formation of $k$th species, J/kg}
\item[$h_k$]{specific enthalpy of $k$th species excluding heat of formation, J/kg}
\item[$h_{\rm e}$]{electron specific enthalpy, J/kg}
\item[$\vec{J}$]{current density vector, A/m$^2$}
\item[$\vec{J}_{\rm e}$]{electron current density vector, A/m$^2$}
\item[$k_{\rm B}$]{Boltzmann constant, J/K}
\item[$k_{\rm e-v}$]{effective electron cooling reaction rate, $\rm m^3/s$}
\item[$k_{\rm v-e}$]{effective electron heating reaction rate, $\rm m^3/s$}
\item[$k_{n}$]{reaction rate of nitrogen vib. excitation to $n$th level, $\rm m^3/s$}
\item[$k_{n}^{\rm inv}$]{reaction rate of nitrogen vib. de-excitation from $n$th level to ground state, $\rm m^3/s$}
\item[$k_{kl}$]{electron impact reaction rate of $l$th process on $k$th neutral species, $\rm m^3/s$}
\item[$k_{{\rm N_2}l}$]{electron impact reaction rate of $l$th process of nitrogen, $\rm m^3/s$}
\item[$m_k$]{particle mass of $k$th species, kg}
\item[$m_{\rm e}$]{electron mass, kg}
\item[$m_{\rm n}$]{mass of a neutral particle, kg}
\item[$m_{\rm N_2}$]{particle mass of nitrogen, kg}
\item[$\rm M_{\infty}$]{freestream Mach number}
\item[$n$]{nitrogen vibrational level}
\item[$N_k$]{number density of $k$th species, $\rm m^{-3}$}
\item[$N_{\rm N_2}$]{number density of nitrogen, $\rm m^{-3}$}
\item[$N_{{\rm N_2}(v)}$]{sum of the number densities of all nitrogen vibrationally excited states, $\rm m^{-3}$}
\item[$N_{\rm e}$]{number density of electrons, $\rm m^{-3}$}
\item[$N$]{number density of gas mixture, $\rm m^{-3}$}
\item[$P_k$]{partial pressure of $k$th species, Pa}
\item[$P$]{pressure of the gas mixture, $\sum P_k$, Pa}
\item[$Q_{\rm e-t} $]{electron cooling-heating rate per unit volume due to elastic collisions, W/m$^3$ }
\item[$Q_{\rm e-v} $]{electron cooling rate per unit volume due to electron-vibrational collisions, W/m$^3$ }
\item[$Q_{\rm v-e} $]{electron heating rate per unit volume due to electron-vibrational collisions, W/m$^3$ }
\item[$Q_{\rm e-i}$]{electron cooling rate per unit volume to all inelastic collisions, W/m$^3$ }
\item[$\overline{{q}_{\rm e}}$]{electron thermal speed, m/s}
\item[$q$]{flow speed, m/s}
\item[$R_{\rm N_2}$]{gas constant of nitrogen, $\rm J/(kg \cdot K)$}
\item[$R$]{leading edge cap radius, m}
\item[$s_k$]{species sign according to charge}
\item[$T$]{bulk gas temperature, K}
\item[$T_{\rm e}$]{electron temperature, K}
\item[$T_{\rm ref}$]{reference temperature in electron swarm experiments, K}
\item[$T_{\rm v}$]{vibrational temperature, K}
\item[$t$]{time, s}
\item[$V_{i}$]{bulk flow velocity component, m/s}
\item[$V_{i}^k$]{component of $k$th species velocity, m/s}
\item[$\vec{V}_{\rm e}$]{electron velocity vector, m/s}
\item[$W_{\rm e}$]{chemical source term of electrons, $\rm kg/(m^3\cdot s)$ }
\item[$W_k$]{chemical source term of $k$th species, $\rm kg/(m^3\cdot s)$ }
\item[$W_{\rm N_2}$]{chemical source term of nitrogen, $\rm kg/(m^3\cdot s)$ }
\item[$w_k$]{mass fraction of $k$th species}
\item[$w_{\rm N_2}$]{nitrogen mass fraction}
\item[$x_{i}$]{Cartesian coordinate, m}
\item[$x$]{axial coordinate, m}
  \end{nomenclaturelist}

  \begin{nomenclaturelist}{Greek symbols}
\item[$\beta_{k}^{\rm n}$]{parameter equal to 1 when $k$th species is neutral, 0 otherwise}
\item[$\beta_{k}^{\rm i}$]{parameter equal to 1 when $k$th species is ionic, 0 otherwise}
\item[$\epsilon_0$]{permittivity of free space, $\rm C/(V\cdot m)$}
\item[$\epsilon_r$]{relative permittivity}
\item[$\gamma_{\rm e}$]{electron secondary emission coefficient}
\item[$\kappa_{\rm e}$]{electron thermal conductivity, $\rm W/(m\cdot K)$}
\item[$\kappa_{\rm i}$]{ion thermal conductivity, $\rm W/(m\cdot K)$}
\item[$\kappa_{\rm n}$]{neutrals thermal conductivity, $\rm W/(m\cdot K)$}
\item[$\kappa_{\rm v}$]{nitrogen vibrational thermal conductivity, $\rm W/(m\cdot K)$}
\item[$\ln \Lambda$]{Coulomb logarithm}
\item[$\mu_k$]{mobility of $k$th charged species,$\rm m^2/(V\cdot s)$}
\item[$(\mu_{\rm e}^\star)_k$]{reduced electron mobility in $k$th neutral species, $(\mu_{\rm e}N)_k$, $\rm m^{-1}V^{-1}s^{-1}$}
\item[$\mu_{\rm e}$]{electron mobility, $\rm m^2/(V\cdot s)$}
\item[$\rho_{k}$]{partial mass density of $k$th species, kg/m$^3$}
\item[$\rho_c$]{net charge density, C/m$^3$}
\item[$\rho_{\rm e}$]{electron mass density, kg/m$^3$}
\item[$\rho_{\rm N_2}$]{nitrogen mass density, kg/m$^3$}
\item[$\rho$]{mass density of the mixture, $\sum_k \rho_k$, kg/m$^3$}
\item[$\eta$]{viscosity, Pa$\cdot$s}
\item[$\phi$]{electric field potential, V}
\item[$\tau_{ji}$]{viscous stress tensor, Pa}
\item[$\tau_{\rm vt}$]{vibrational-translational relaxation time, s}
\item[$\theta_{\rm v}$]{characteristic nitrogen vibration temperature, K}
\item[$\theta_{\rm e}$]{closure coefficient from detailed balancing in electron heating rate, K}
\item[$\nu_k$]{mass diffusion coefficient of $k$th species, $\rm kg/(m\cdot s)$}
\item[$\nu_{\rm N_2}$]{mass diffusion coefficient of nitrogen, $\rm kg/(m\cdot s)$}
\item[$\chi_{\rm e}$]{plasma ionization fraction}
\item[$\zeta_{\rm e}$]{mean fraction of electron kinetic energy lost in one collision}
\item[$\zeta_{\rm v}$]{fraction of electron energy consumed in the excitation of the nitrogen vibrational energy}
  \end{nomenclaturelist}

}

\begin{document}
\maketitle
\makenomenclature

\section{Introduction}

\dropword Electron-vibrational (e-V) coupling, the energy exchange between free electrons and the vibrational modes of molecules, is a critical process of central importance in many applications, including the non-equilibrium plasmas found in hypersonic flight, plasma-assisted combustion (PAC), and laser-induced plasmas (LIP).

In hypersonic flight beyond Mach 10, a plasma layer forms around the vehicle that can interfere with radio communications but also enables various advanced technologies. As explored in early work by \citet{gmrl:1963:musal} and later by \citet{dtic:1992:gregorie}, this includes electromagnetic shielding. More recent applications include plasma antennas, discussed by \citet{ieee:2024:magarotto}; electron transpiration cooling (ETC), detailed by \citet{aip:2017:hanquist} and \citet{aiaapaper:2021:parent}; and magnetohydrodynamic (MHD) systems, which have been the focus of numerous studies such as the ones by \cite{aiaaconf:2004:macheret}, \cite{aiaaconf:2022:moses}, and \cite{jtht:2023:parent,aiaa:2025:parent}. Harnessing these applications requires the accurate prediction of plasma properties like electrical conductivity and plasma frequency, which are fundamentally linked to the electron temperature ($T_{\rm e}$). The collision cross sections needed to determine electrical conductivity, for instance, are a function of electron temperature. Furthermore, electrical conductivity depends on the plasma density, which is also a function of $T_{\rm e}$ through processes like electron-ion recombination and ambipolar diffusion to surfaces as noted in \cite{pf:2022:parent}. Plasma density, in turn, governs the plasma frequency, which is crucial for electromagnetic interactions.

Accurate modeling of $T_{\rm e}$ requires a precise description of the energy exchange between electrons and heavy particles via elastic and inelastic collisions. In the non-equilibrium conditions typical of hypersonic flows where nitrogen is not fully dissociated, the dominant energy exchange mechanism involves inelastic collisions with vibrationally excited nitrogen molecules (e-V coupling). Characteristically for non-equilibrium re-entry flows, $T_{\rm e}$ is lower than both the bulk gas and nitrogen vibrational temperatures ($T_{\rm v}$), making electron \emph{heating} through superelastic collisions the controlling process. However, direct experimental or molecular dynamics data for these heating rates are lacking, forcing models to infer them from the corresponding electron \emph{cooling} rates.

{
Plasma-assisted combustion is another field where e-V coupling is paramount. As summarized in \cite{pecs:2015:ju}, high-energy electrons from the discharge initiate parallel kinetic and thermal enhancement pathways. The kinetic pathway produces radicals and a population of excited species. The thermal pathway has two components: fast heating from the rapid quenching of the electronically excited states as observed by \cite{apl:2011:bak}, and slow heating from the gradual vibrational-translational (V-T) relaxation of the vibrationally-excited nitrogen that was created via e-V energy transfer. Detailed kinetic mechanisms, such as the ones by \cite{rsta:2015:adamovich} and \cite{jap:2023:chen}, incorporate state-specific N$_2$ vibrational energy transfer to model this slow heating pathway. For turbulent simulations, phenomenological models were developed; for instance, \cite{cf:2016:castela} proposed that 45\% of the discharge energy goes into N$_2$ vibrational excitation, a model also used by \cite{cf:2024:taneja} for large eddy simulations of plasma assisted ignition. However, \cite{cf:2025:dijoud} showed this energy partition is not fixed and varies with fuel-oxidizer composition, temperature, and pulse shape. Therefore, a predictive model for V-T energy transfer is crucial for accurately simulating PAC across different operational regimes. The accuracy of this model, in turn, depends on e-V coupling, which sets the electron temperature that governs the production of vibrationally excited states.

Another area where electron-vibrational (e-V) coupling is critical is the study of laser-induced plasmas. These are used for various applications ranging from non-intrusive flow diagnostics as described by \cite{jap:2019:peters} to long-distance microwave guiding explored by \cite{jap:2010:shneider}. Simulating these phenomena requires detailed kinetic models that resolve the evolution of separate translational, vibrational, and electron temperatures. In models developed by \cite{jap:2019:peters} and \cite{jap:2023:pokharel}, electron-impact vibrational excitation of N$_2$ is a dominant energy loss mechanism for electrons. The reverse process, vibrational heating via superelastic collisions, becomes a critical energy source for electrons at later times, slowing the decay of $T_{\rm e}$. This e-V energy exchange is paramount, as the electron temperature directly controls the rates of electron attachment and dissociative recombination, which govern the plasma's overall lifetime. For instance, as shown in \cite{jap:2010:shneider}, a secondary laser pulse can be used to intentionally heat electrons and suppress these loss mechanisms, thereby extending the lifetime of a plasma waveguide.}

Several approaches have been employed to model vibrational-electron heating. Early models, used in works like \cite{jap:2010:shneider}, assumed the heating rate scales with the cooling rate by a factor of $T_{\rm v}/T_{\rm e}$. Similarly, \cite{jtht:2012:kim} and \cite{jtht:2013:farbar} proposed scaling based on the ratio of vibrational energies, $e_{\rm v}(T_{\rm v})/e_{\rm v}(T_{\rm e})$. While these formulations correctly drive the system towards $T_{\rm e}=T_{\rm v}$ at equilibrium, they lack a rigorous derivation from fundamental principles, limiting their reliability in strong thermal non-equilibrium. More recently, attempts by \cite{jap:2019:peters} and \cite{jap:2023:pokharel}  utilized the principle of detailed balance to connect heating and cooling rates. Although seemingly rigorous, this specific application leads to a significant flaw: it fails to guarantee convergence to $T_{\rm e} = T_{\rm v}$ at equilibrium, thus violating thermodynamic consistency.

This work introduces a novel, thermodynamically consistent model for electron heating. Our approach derives the heating rates from the corresponding cooling rates by invoking the equilibrium constant relationship between forward and reverse reactions. This relationship is expressed in terms of the vibrational temperature using an effective activation energy and assuming a Boltzmann distribution for the vibrationally excited states. This method essentially satisfies detailed balance while ensuring the net energy transfer rate depends correctly on both $T_{\rm e}$ and $T_{\rm v}$. Critically, unlike the models of \cite{jap:2023:pokharel} and of \cite{jap:2019:peters}, our formulation guarantees the correct physical behavior: it drives the electron temperature towards the vibrational temperature  as equilibrium is approached and yields zero net energy exchange precisely when $T_{\rm e} = T_{\rm v}$. This provides a robust and physically sound closure for the electron energy equation.

The proposed model's validity and predictive capability are assessed through simulations compared against experimental data from diverse test cases: (i) plasma flows in uniform electric fields, (ii) the RAM-C-II {and OREX re-entry flight experiments,} and (iii) a test case to evaluate the performance of previous and proposed electron heating models towards thermal equilibrium.

\section{Transport Equations}

For non-neutral or quasi-neutral plasmas, the motion of the charged species with
respect to the bulk can be found through the ‘drift-diffusion’ plasma model. A very good approximation is to neglect the inertia terms of electrons and ions as well as shear stresses or forces due to collisions between charged species, since these are negligible compared to forces in electron-neutral or ion-neutral collisions when the plasma is \textit{weakly ionized}. Thus, the species velocities can be shown to be, as outlined in \cite{book:2022:parent}:
\begin{equation}
  V^{k}_i = \left\{
  \begin{array}{ll}\mfd
  V_i+s_k \mu_k  {E}_i
             -  \frac{\mu_k}{|C_k| N_k} \frac{\partial P_k}{\partial x_i} & \textrm{for charged species} \alb\mfd
  V_i - \frac{\nu_k}{\rho_k} \frac{\partial w_k}{\partial x_i} & \textrm{for neutral species}
  \end{array}
  \right.
\label{eqn:Vk}
\end{equation}
where $V^{k}_i$ is the species velocity, $V_i$ is the bulk velocity of the plasma, $E$ is the electric field, $\mu_k$ is the mobility, $N_k$ is the species number density and $P_k$ is the partial pressure. As well, $C_k$ is the elementary charge of single-charge species and $s_k$ is the sign of the charge of the species under consideration (-1 for electrons or negative ions, +1 for positive ions). For neutral species, the velocity $V^{k}_i$ is written in terms of the diffusion term with respect to the bulk, where $\nu_k$ is the mass diffusion coefficient of neutral species $k$, $\rho_k$ is the mass density and $w_k$ is the mass fraction of species $k$. For each $k$th  species, either charged or neutral, we can express the mass conservation equation as follows:
\begin{equation}
\frac{\partial}{\partial t} \rho_k + \sum_i \frac{\partial }{\partial x_i}\rho_{k} V_i^{k} = W_{k}  
\label{eqn:masstransport}
\end{equation}
The momentum equation for the bulk flow corresponds to the Navier-Stokes equations including the electric field body force:
\begin{equation}
  \rho \frac{\partial V_i }{\partial t}+ \sum_{j=1}^3 \rho V_j \frac{\partial V_i}{\partial x_j}
=
-\frac{\partial P}{\partial x_i} 
+ \sum_{j=1}^3 \frac{\partial \tau_{ji}}{\partial x_j}
+ \rho_{\rm c}{E}_i
\label{eqn:momentumtransport}
\end{equation}
where $P=\sum_k P_k$ is the sum of all species partial pressures, $\tau_{ji}$ is the Navier-Stokes viscous stress tensor and $\rho_{\rm c}=\sum_k C_k N_k$ corresponds to the net charge density.

Within the framework of a non-equilibrium, three-temperature model, additional energy transport equations are required to obtain the translational, vibrational and electron temperatures. Because the vibrational energy coupling of diatomic air molecules such as $\rm O_2$ and $\rm NO$ is {here assumed} weak compared to that of molecular nitrogen, only the latter is considered in the vibrational energy transport. The nitrogen vibrational energy transport equation corresponds to:
\begin{equation}
 \begin{array}{l}
  \mfd\frac{\partial}{\partial t} \rho_{\rm N_2} e_{\rm v}
     + \sum_{j=1}^{3} \frac{\partial }{\partial x_j}
       \rho_{\rm N_2} V_j e_{\rm v}
     - \sum_{j=1}^{3} \frac{\partial }{\partial x_j} \left(
            \kappa_{\rm v}  \frac{\partial T_{\rm v}}{\partial x_j}\right)\alb\mfd
     - \sum_{j=1}^{3} \frac{\partial }{\partial x_j} \left(
            e_{\rm v} \nu_{\rm N_2}  \frac{\partial w_{\rm N_2}}{\partial x_j}\right)
 = 
Q_{\rm e-v}-Q_{\rm v-e}   + \frac{\rho_{\rm N_2}}{\tau_{\rm vt}}\left( e_{\rm v}^0 -e_{\rm v} \right) + W_{\rm N_2} e_{\rm v}
\end{array}
\label{eqn:vibrationalenergytransport}
\end{equation}
where $e_{\rm v}$ , $e_{\rm v}^0$ and $\kappa_{\rm v}$ are the nitrogen vibrational energy, vibrational energy in equilibrium and the nitrogen vibrational thermal conductivity, respectively. As well, $Q_{\rm e-v}-Q_{\rm v-e}$ is the net gain of energy from the electrons to the nitrogen vibrational energy modes (which will be outlined later), while $\tau_{\rm vt}$ is the {V-T} relaxation time taken from  \cite{aiaa:2001:macheret}. 

The electron energy transport equation is derived from the first law of thermodynamics applied to electrons as done in \cite{book:1991:raizer} neglecting electron inertia terms, and takes on the following form:
\begin{align}
 \frac{\partial }{\partial t}\rho_{\rm e} e_{\rm e} &+ \sum_{j=1}^3  \frac{\partial }{\partial x_j} \left( \rho_{\rm e} V^{\rm e}_j h_{\rm e}  - \kappa_{\rm e} \frac{\partial T_{\rm e}}{\partial x_j}  
\right)\nonumber \alb
&= 
 W_{\rm e} e_{\rm e}
+   C_{\rm e} N_{\rm e} \vec{E} \cdot \vec{V}_{\rm e}
 -Q_{\rm e-t}-Q_{\rm e-i}+Q_{\rm v-e}
\label{eqn:electronenergytransport}
\end{align}
where $e_{\rm e}$, $h_{\rm e}$ and $\kappa_{\rm e}$ are the electron specific energy, electron specific enthalpy and electron thermal conductivity respectively. In the latter {$Q_{\rm e-i}$ is the electron cooling due to all inelastic collision processes, $Q_{\rm v-e}$ is the electron heating due to electron-vibrational collisions, while} $Q_{\rm e-t}$ is the electron cooling-heating due to elastic collisions, which we take from \cite{pf:2024:parent}:
\begin{align}
Q_{\rm e-t}
&= 
   \underbrace{\sum_k  \frac{3 \beta_k^{\rm n} k_{\rm B} |C_{\rm e}| N_{\rm e} N_k  (T_{\rm e}-T)}{ m_k (\mu_{\rm e}N)_k}}_{\rm elastic~cooling-heating~to~neutrals}\nonumber\\
 &+ \underbrace{\sum_k \beta_k^{\rm i}  N_{\rm e} N_k  (T_{\rm e}-T )  \frac{6 k_{\rm B} C_{\rm i}^2 C_{\rm e}^2 \ln \Lambda}{ \pi^3 \epsilon_0^2 m_{\rm e} m_k \overline{q_{\rm e}}^3}}_{\rm elastic~cooling-heating~to~ions} 
\label{eqn:Qelastic}
\end{align}
with $\ln \Lambda$ the recommended Coulomb logarithm which can be found in \cite[page 34]{nrl:2002:huba} and $(\mu_{\rm e} N)_k\equiv (\mu_{\rm e}^\star)_k$ corresponds to the reduced electron mobility of the $k$th neutral species.

To arrive at an expression for the total inelastic electron {cooling} rate {(due to electron-vibrational, electron-rotational, electron-ionization, and other inelastic electron cooling processes)}, let us first start with the rate of electron energy loss in inelastic collisions written in standard form as in \cite{jap:2019:peters} or \cite{jap:2023:pokharel}:
\begin{equation}
Q_{\rm e-i}  
=   \sum_k   \beta_k^{\rm n} N_{\rm e} N_k  \sum_l k_{kl} \mathcal{E}_{kl}
\label{eqn:Qinelastic_standard}
\end{equation}
where $l$ denotes an electron impact process, $k_{kl}$ is the rate coefficient of the $l$th electron impact process acting on the $k$th neutral species, $\mathcal{E}_{kl}$ is the activation energy of the $l$th electron impact process of the $k$th species and $\beta_k^{\rm n}$ is 1 when the $k$th species is a neutral and to 0 otherwise. As was shown by \cite{pf:2024:parent}, the  electron cooling can be expressed as a function of the $k$th species reduced electric fields $E^\star_k$ and reduced mobilities $\left(\mu_{\rm e}^\star\right)_k$ without any loss of generality as follows:
\begin{equation}
  \sum_l k_{kl} \mathcal{E}_{kl}  
=  |C_{\rm e}|  \left(\mu_{\rm e}^\star\right)_k \left( (E^\star_k)^2 - \frac{3  k_{\rm B}    (T_{\rm e}-T_{\rm ref})}{ m_{k} (\mu_{\rm e}^\star)^2_k}\right) 
\label{eqn:suminelasticenergies}
\end{equation}
We can then substitute the latter in the former to obtain the total electron cooling {due to inelastic collisions}:
\begin{equation}
Q_{\rm e-i}  
=  \sum_k \beta_k^{\rm n} |C_{\rm e}| N_{\rm e} N_k (\mu_{\rm e}^\star)_k \left( (E^\star_k)^2 -  \frac{3  k_{\rm B}    (T_{\rm e}-T_{\rm ref})}{ m_{k} (\mu_{\rm e}^\star)^2_k} \right) 
\label{eqn:Qinelastic_cooling}
\end{equation}
Expressing electron cooling function of reduced electric fields and reduced mobilities instead of rates for all cooling processes is more accurate because the reaction rates obtained using cross-sectional data and BOLSIG+ have considerable error especially when the electron temperature is less than 1~eV as is often the case in hypersonic flows. Indeed, the reduced electric field and reduced mobility can be obtained with high accuracy even at low electron temperature using swarm experiments.

The translational temperature $T$ is determined from the total energy equation, obtained by summing all energy transport equations:
\begin{equation}
\begin{array}{l}\mfd
 \frac{\partial }{\partial t}\rho e_{\rm t}
+ \sum_{j=1}^3  \frac{\partial }{\partial x_j} V_j \left(\rho  e_{\rm t} +  P \right)
 \alb\mfd
- \sum_{j=1}^3  \frac{\partial }{\partial x_j} \left(
   \nu_{\rm N_2} e_{\rm v}\frac{\partial w_{\rm N_2}}{\partial x_j} 
  + \sum_{k=1}^{\rm n_s}  \rho_k (V^k_j-V_j) {(h_k+h_k^\circ)}
\right)
 \alb\mfd
-\sum_{i=1}^{3}\frac{\partial }{\partial x_i}\left((\kappa_{\rm n}+\kappa_{\rm i}) \frac{\partial T}{\partial x_i} 
+ \kappa_{\rm v} \frac{\partial T_{\rm v}}{\partial x_i} +\kappa_{\rm e} \frac{\partial T_{\rm e}}{\partial x_i}\right)
 \alb\mfd
=
 \sum_{i=1}^3 \sum_{j=1}^3  \frac{\partial }{\partial x_j} \tau_{ji} V_i
+ \vec{E}\cdot\vec{J}
\end{array}
\label{eqn:totalenergytransport}
\end{equation}
where $h_k^\circ$ stands for the heat of formation while the sum $(h_k+h_k^\circ)$ represents the enthalpy of the $k$th species including calorically-imperfect effects as well as the heat of formation, obtained from the NASA Glenn high temperature enthalpy polynomials by \cite{nasa:2002:mcbride}. As well, $\vec{J} \equiv \sum_{k}  C_k N_k  \vec{V}_k$ is the current density vector and $e_{\rm t}$ is the total specific energy. To close the system of equations, the electric field $\vec{E}$ is here found as the negative of the gradient of the electric potential obtained from Gauss's law:
\begin{equation} 
\sum_{i=1}^3 \frac{\partial }{\partial x_i}\left(\epsilon_r \frac{\partial \phi}{\partial x_j}\right)=-\frac{\rho_{\rm c}}{\epsilon_0}   \label{eqn:gauss_potential} 
\end{equation}
with $\phi$, $\rho_{\rm c}$, $\epsilon_0$, and $\epsilon_{\rm r}$   the electric field potential, the net charge density, the permittivity of free space, and the relative permittivity, respectively. 

At solid surfaces, we assume no surface catalysis for neutral species. For charged species it is assumed that electrons and ions recombine with a probability of 1 and secondary electron emission is modeled by the effective secondary electron emission coefficient $\gamma_{\rm e}$ which is set to 0.1 for all test cases. 

Chemical kinetics are obtained for 11 air species $\rm N_2$, $\rm O_2$, $\rm NO$, $\rm N$, $\rm O$, $\rm N_2^+$, $\rm O_2^+$, $\rm NO^+$, $\rm N^+$, $\rm O^+$ and $\rm e^-$ using a Park-like solver by \cite{ijhmt:2021:kim} with corrected reaction rates outlined in \cite{pf:2024:parent}.

Unless otherwise indicated, the mobilities, viscosity, thermal conductivity, and other transport coefficients are taken from \cite{nasa:1990:gupta}. The latter is corrected following \cite{jtht:2023:parent} to improve the agreement with experimental air data on the basis of electrical and thermal conductivity.

\section{Proposed Vibrational-Electron Heating Model}

\begin{figure}[t]
    \centering
     \includegraphics[width=0.40\textwidth]{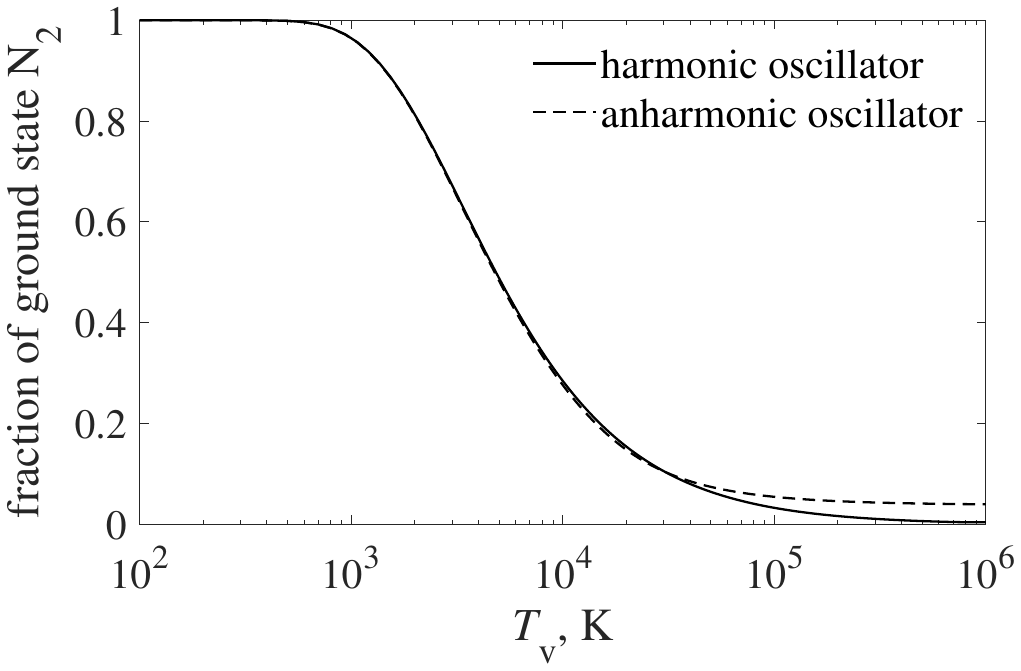}
     \figurecaption{Effect of the assumption of harmonic oscillator on the fraction of ground state nitrogen.}
     \label{fig:fraction_v0_N2}
\end{figure}

We here outline a derivation from basic principles of a novel electron heating model that both satisfies the detailed balance principle and that is thermodynamically consistent. This is here accomplished by first defining an \emph{effective} activation energy as the average activation energy of all vibrationally excited states, as follows:
{
\begin{equation}
 \mathcal{E}_{\rm eff}
\equiv
\frac{\textrm{vibrational energy excluding zero-point energy}}{\textrm{number of vibrationally excited molecules}}
\end{equation}
Noting that $e_{\rm v}$ is the average nitrogen vibrational energy per unit mass 
(excluding the zero-point energy) and defining $N_{{\rm N_2}(v)}$ as the sum of the number densities of all $\rm N_2$ vibrationally excited states (excluding the ground state), the latter can also be written as:}
\begin{equation}
 \mathcal{E}_{\rm eff}
=
\frac{\rho_{\rm N_2} e_{\rm v}}{N_{{\rm N_2}(v)}}
\label{eqn:eff_act_energy}
\end{equation}
We can substitute the partial density of nitrogen $\rho_{\rm N_2}$ by $N_{\rm N_2} m_{\rm N_2}$ with $N_{\rm N_2}$ being the number density of all nitrogen molecules and $m_{\rm N_2}$ the mass of one nitrogen molecule. It is emphasized that $N_{\rm N_2}$ is not the number density of the ground state of $\rm N_2$. Rather, it is the sum of the number densities of all $\rm N_2$ states, including the ground state and all excited states. Then, using the symbol $f_v$ to denote the fraction of N$_2$ within the $v$ vibrational state, Eq.~(\ref{eqn:eff_act_energy}) can be rewritten to:
\begin{equation}
  \mathcal{E}_{\rm eff} \sum_{v=1}^\infty f_v
=
m_{\rm N_2} e_{\rm v}
\label{eqn:eff_act_energy_3}
\end{equation}
Noting that the sum of all $f_v$s including the ground state $v=0$ must equal 1, it follows that:
\begin{equation}
\left(1- f_{v=0} \right) \mathcal{E}_{\rm eff}
=
m_{\rm N_2} e_{\rm v}
\label{eqn:eff_act_energy_4}
\end{equation}
We can obtain $f_v$ from the Boltzmann distribution described in \cite[pp. 518, 578]{book:1989:anderson} and \cite[pp. 111, 418]{book:2012:capitelli} given $T_{\rm v}$:
\begin{equation}
f_v = 
\exp\left(\frac{-\mathcal{E}_v }{k_{\rm B} T_{\rm v}}\right)
\left/
\sum_{n=0}^{\infty} \exp\left(\frac{-\mathcal{E}_{n}}{ k_{\rm B} T_{\rm v}}\right) 
\right.
\label{eqn:vib_frac_full}
\end{equation}
where $\mathcal{E}_{v}$ is the energy of vibrational level $v$. The assumption of an equilibrium distribution of nitrogen vibrational levels is valid as long as vibrational-vibrational and vibrational-translational relaxation processes are sufficiently fast. Such was already assumed in the previous section where a  single vibrational temperature describes all vibrational energies of the nitrogen molecule. We can further assume an harmonic oscillator model without introducing significant error. Then, as outlined in \cite[page 436]{book:1989:anderson}~and~\cite[pp. 134, 135]{book:1965:vincenti}, the $n$th vibrational state activation energy can be approximated as $\mathcal{E}_n= \theta_{\rm v}  k_{\rm B} n$. In the latter, $\theta_{\rm v}$ is the  characteristic vibration temperature. For nitrogen such corresponds to 3353~K, as suggested in \cite{book:1962:barrow}. Therefore, the fraction of the ground state corresponds to:
\begin{equation}
f_{v=0}=\frac{1}{\mfd \sum_{n=0}^\infty \exp\left(-\frac{n \theta_{\rm v}}{T_{\rm v}}\right)} 
\end{equation}
This infinite geometric series can be shown to become simply $(1-\exp(-\theta_{\rm v}/T_{\rm v}))^{-1}$ following \cite[page 438]{book:1989:anderson}. Then, it follows that the fraction of the ground state of nitrogen becomes simply:
\begin{equation}
f_{v=0} = 1 - \exp({-\theta_{\rm v} / T_{\rm v}}) 
\label{eqn:vib_frac_harmonic}
\end{equation}
In Fig.~\ref{fig:fraction_v0_N2}, a comparison is outlined between the latter approximate expression assuming a harmonic oscillator model including all energy levels, and the anharmonic one outlined in Eq.~(\ref{eqn:vib_frac_full}) summing up to 25 levels. The anharmonic energy levels are taken from \cite{psst:2014:laporta}. It can be seen that the error is less than 2\% when the vibrational temperature is 30,000 K or less.
We can substitute Eq.~(\ref{eqn:vib_frac_harmonic}) in Eq.~(\ref{eqn:eff_act_energy_4}) to obtain a much simpler expression for the effective activation energy:
\begin{equation}
\mathcal{E}_{\rm eff}
=
\frac{m_{\rm N_2} e_{\rm v}}{\exp(- \theta_{\rm v} / T_{\rm v})}
\end{equation}
{We can obtain the average vibrational energy $e_v$ excluding its zero-point energy from the Planck distribution, which is derived from the Bose-Einstein statistics assuming an harmonic oscillator model:}
\begin{equation}
e_{\rm v} = \dfrac{R_{\rm N_2}\theta_{\rm v}}{\exp(\theta_{\rm v}/T_{\rm v})-1}
\label{eqn:ev_Tv}
\end{equation}
where $R_{\rm N_2}$ is the specific gas constant for molecular nitrogen. Substituting the latter equation in the former and noting that $R_{\rm N_2}=k_{\rm B}/m_{\rm N_2}$, the effective activation energy simplifies to:
\begin{equation}
\mathcal{E}_{\rm eff}
= \dfrac{k_{\rm B}\theta_{\rm v}}{1-\exp(-\theta_{\rm v}/T_{\rm v})}
\label{eqn:Eeff_Tv}
\end{equation}
Let us keep the latter on hold. The second step in our derivation process consists of defining an \textit{effective} electron cooling reaction rate $k_{\rm e-v}$ such that, when multiplied by the effective activation energy, it yields the {electron-vibrational} energy loss due to inelastic collisions with the nitrogen molecules. Thus the effective electron cooling rate is equal to:
\begin{equation}
  k_{\rm e-v}   
\equiv \frac{1}{\mathcal{E}_{\rm eff}} \sum_{l}^{\rm vib} k_{{\rm N_2}l} \mathcal{E}_{{\rm N_2}l}
\label{eqn:kcool}
\end{equation}
where $\mathcal{E}_{{\rm N_2}l}$ and  and $k_{{\rm N_2}l}$ are the activation energy and the reaction rate associated with the $l$th electron-vibrational cooling inelastic process of the N$_2$ molecule, respectively. 

{Applying the principle of detailed balance to the macroscopic forward (cooling) and reverse (heating) rates yields an equilibrium constant, which allows us to express the effective reaction rate for the heating process as follows:}
\begin{equation}
k_{\rm v-e}=k_{\rm e-v} \exp \left(\frac{\theta_{\rm e}}{T_{\rm e}} \right)
\label{eqn:kheat}
\end{equation}
where $\theta_{\rm e}$ is a closure coefficient proportional to the activation energy which will be determined later.  

{Because inelastic electron heating is due to loss of energy of a vibrationally excited state due to the collision between the latter and electrons, it can be written as the product between the effective electron heating reaction rate $k_{\rm v-e}$, the average activation energy, the number  of vibrationally excited molecules, and the number of electrons:}
\begin{equation}
Q_{\rm v-e}=N_{\rm e} N_{{\rm N_2}(v)} k_{\rm v-e} \mathcal{E}_{\rm eff}
\end{equation}
where $N_{{\rm N_2}(v)}$ is the sum of the number densities of all $\rm  N_2$ vibrationally excited states. But recall that the definition of the effective activation energy outlined in Eq.~(\ref{eqn:eff_act_energy}) entails that $N_{{\rm N_2}(v)} \mathcal{E}_{\rm eff}=\rho_{\rm N_2} e_{\rm v}$. Further, we can substitute $k_{\rm v-e}$ from Eq.~(\ref{eqn:kheat}) to obtain:
\begin{equation}
Q_{\rm v-e}=N_{\rm e} \rho_{\rm N_2} e_{\rm v} k_{\rm e-v} \exp \left(\frac{\theta_{\rm e}}{T_{\rm e}} \right)
\end{equation}
Further substituting $k_{\rm e-v}$ from Eq.~(\ref{eqn:kcool}), it follows that:
\begin{equation}
Q_{\rm v-e}=N_{\rm e} \rho_{\rm N_2}  \frac{e_{\rm v}}{\mathcal{E}_{\rm eff}  } 
\exp \left(\frac{\theta_{\rm e}}{T_{\rm e}} \right)
\sum_{l}^{\rm vib} k_{{\rm N_2}l} \mathcal{E}_{{\rm N_2}l} 
\end{equation}
Finally, we can rewrite $\mathcal{E}_{\rm eff}$ and $e_{\rm v}$ in terms of $T_{\rm v}$ using Eqs. (\ref{eqn:ev_Tv}) and (\ref{eqn:Eeff_Tv}) to arrive at:
{
\begin{equation}
Q_{\rm v-e}=N_{\rm e} \rho_{\rm N_2} \dfrac{R_{\rm N_2}}{k_{\rm B}}
\dfrac{1-\exp(-\theta_{\rm v}/T_{\rm v})}{\exp(\theta_{\rm v}/T_{\rm v})-1} 
\exp \left(\frac{\theta_{\rm e}}{T_{\rm e}} \right)
\sum_{l}^{\rm vib} k_{{\rm N_2}l} \mathcal{E}_{{\rm N_2}l} 
\end{equation}
Noting that $\rho_{\rm N_2}=N_{\rm N_2} m_{\rm N_2}$, that $R_{\rm N_2}=k_{\rm B}/m_{\rm N_2}$, and after multiplying both the numerator and denominator by $\exp(\theta_{\rm v}/T_{\rm v})$ and simplifying, the following is obtained:}
\begin{align}
Q_{\rm v-e}
&= 
 \exp \left(\dfrac{\theta_{\rm e}}{T_{\rm e}}-\dfrac{\theta_{\rm v}}{T_{\rm v}}\right) 
 \underbrace{ N_{\rm e} N_{\rm N_2} \sum_{l}^{\rm vib} k_{{\rm N_2}l} \mathcal{E}_{{\rm N_2}l} }_{\rm rate~of~N_2~e-V~cooling}
\label{eqn:Qe_inelastic_heating}
\end{align}
Noting that the last terms on the RHS correspond to the rate of {electron-vibrational} cooling by the N$_2$ molecule, we can say:
\begin{align}
\frac{Q_{\rm v-e}}{Q_{\rm e-v}}
&= 
 \exp \left(\dfrac{\theta_{\rm e}}{T_{\rm e}}-\dfrac{\theta_{\rm v}}{T_{\rm v}}\right) 
\label{eqn:scale_proposed_preliminary}
\end{align}
Recall that $\theta_{\rm e}$ is a closure coefficient which needs to be determined. To ensure that the model is thermodynamically consistent, we will here impose that the ratio between the electron heating and cooling due to inelastic collision is 1 when the electron temperature equals the vibrational temperature. This will lead to a  thermodynamically consistent model by ensuring that there is no net energy exchange in the limit of thermal equilibrium. It is easy to see that such will occur when
\begin{equation}
\theta_{\rm e}=\theta_{\rm v}
\end{equation}
Recalling Eq.~(\ref{eqn:kheat}), and noting that $\theta_{\rm v}=\mathcal{E}_{v=1}/k_{\rm B}$ with $\mathcal{E}_{v=1}$ the activation energy of the first vibrational level of the nitrogen molecule, this choice of $\theta_{\rm e}=\theta_{\rm v}$ to ensure thermodynamic consistency effectively entails that the equilibrium constant is here determined from the electron cooling process which leads to the ground state gaining energy and becoming the first vibrationally excited state {(the fundamental transition), or from another process requiring the same energy transition as the fundamental transition. In the context of vibrationally-excited states, such is commonly referred to as overtone-less processes. Such is an excellent assumption for many plasma flows as will be outlined in the next subsection.} 

\begin{figure}[t]
    \centering
     \includegraphics[width=0.40\textwidth]{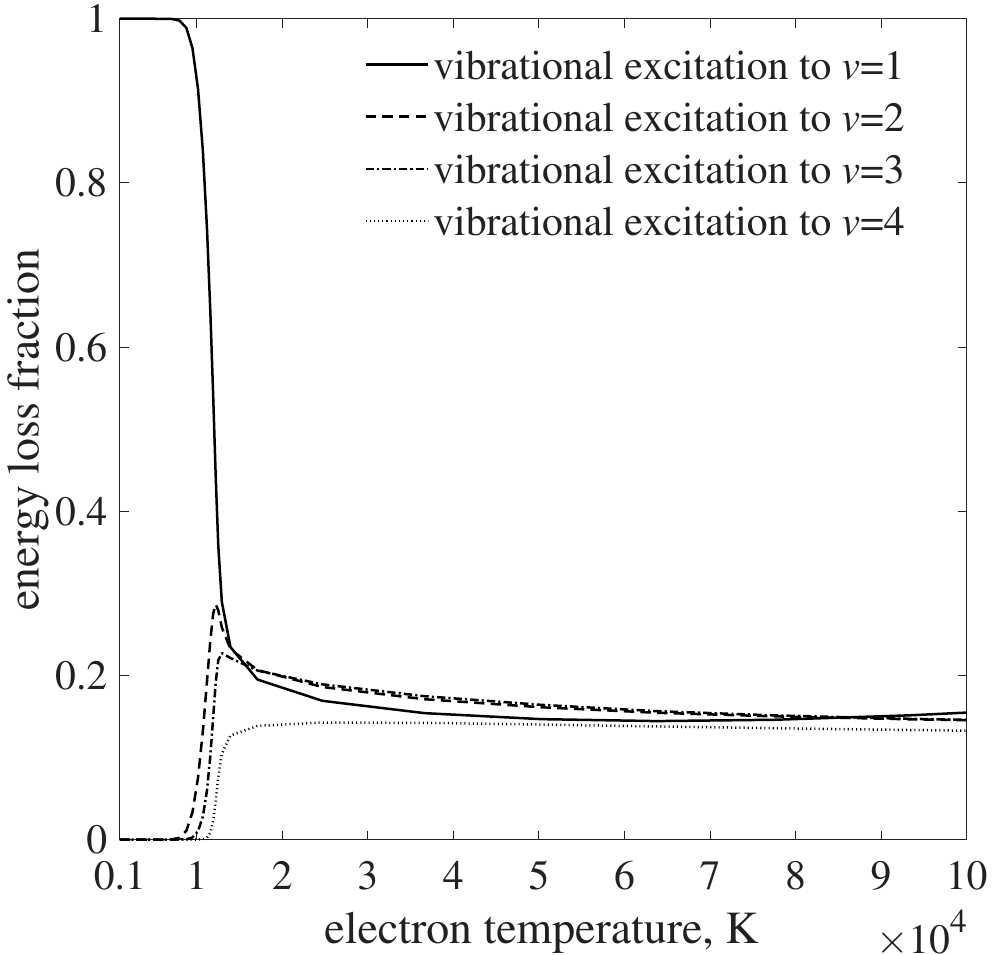}
     \figurecaption{Fractional electron energy loss due to excitation of $\rm N_2$ vibrational levels.}
     \label{fig:sumratesenergy_N2}
\end{figure}

Therefore, although setting $\theta_{\rm e}=\theta_{\rm v}$ does not result in an exact detailed balance process, it does result in a process that is very close to the detailed balance while being thermodynamically consistent. Thus, the electron heating model we recommend is the following:
\begin{align}
\frac{Q_{\rm v-e}}{Q_{\rm e-v}}
&= 
 \exp \left(\dfrac{\theta_{\rm v}}{T_{\rm e}}-\dfrac{\theta_{\rm v}}{T_{\rm v}}\right) 
\label{eqn:scale_proposed}
\end{align}
{
Although the above derivation was done for the nitrogen molecule, it remains valid for other molecules as well as will be discussed below in Section III.G. But before discussing this, let us outline the validity of the 3 main assumptions that were made while deriving the above expression for electron heating: the overtone-less assumption, the Boltzmann distribution assumption, and the detailed balance assumption.

\subsection{Validity of the Overtone-Less Assumption}

Recall that in deriving the previous expression for V-e heating, it was necessary to impose that the equilibrium constant is determined from electron cooling processes which lead to an amount of energy exchange equal to the fundamental transition (i.e., $\theta_{\rm e}=\theta_{\rm v}$). } To assess whether this assumption is sound, we can obtain the fraction of electron cooling due to vibrational excitation of the first level and above. To do this, we compute the rates of inelastic cooling in Eq.~(\ref{eqn:Qe_inelastic_heating}) for vibrational excitation from the ground state to the first four vibrational levels. We then divide each of the latter by the rate of $\rm N_2$ inelastic cooling due to all vibrational excitation processes. As can be seen in Fig.~\ref{fig:sumratesenergy_N2}, assigning $\theta_{\rm e}=\theta_{\rm v}$ is an excellent approximation for plasmas {in which the electron temperature remains below 1~eV} because less than 10\% of the inelastic cooling of the nitrogen molecule is due {to overtones (gain of energy leading to a level skip)} within the electron temperature range 450--10,000~K. { Although the latter only applies to the excitation of the ground state, the transition of the hot bands (vibrationally excited states) to higher levels require almost the same energy transition as the fundamental transition within the harmonic oscillator model, further validating our use of $\theta_{\rm e}=\theta_{\rm v}$. }

Such a range of electron temperature is expected in most hypersonic or reentry flows. Indeed, should the air plasma reach thermal equilibrium, the temperature of the mixture would not exceed 10,000~K unless the flight Mach number is in excess of 33. For flows in non-thermal equilibrium, electrons are weakly coupled to heavy particles, and energy transfer from heavy species to electrons occurs primarily through inelastic collisions, especially with vibrationally excited molecules. Since the vibrational temperature of nitrogen is often significantly lower than the translational temperature and evolves more slowly, the electron temperature tends to equilibrate with the vibrational temperature, keeping it below the latter and typically well under 10,000~K, {thus justifying the overtone-less assumption made herein. }

{In applications such as plasma-assisted combustion (PAC) or laser-induced plasma (LIP), however, the preceding assumption breaks down. Here, discharges or laser pulses can raise the electron temperature to 3–6 eV. In this high-temperature regime, the detailed model from \cite{jap:2019:peters} is necessary to account for overtones. A key deficiency of this model, however, is that it is not thermodynamically consistent at lower electron temperatures, which can lead to unphysical results. This flaw becomes critical during the cool-down phase between pulses. Accurately modeling the V-e energy exchange as $T_{\rm e}$ drops below 10,000 K is crucial because the electron temperature governs electron-ion recombination. This rate, in turn, sets the initial plasma density for the subsequent pulse—a key parameter, since a higher density from pre-ionization lowers the energy threshold for the next breakdown. Therefore, the recommended strategy for simulating PAC or LIP is to use the model from \cite{jap:2019:peters} when $T_{\rm e}>1$~eV but switch to the thermodynamically-consistent formulation in Eq.~(\ref{eqn:scale_proposed}) when the electron temperature is lower.}

{
\subsection{Validity of the Boltzmann Distribution Assumption}

A central assumption of the proposed model is that the vibrational levels of N$_2$ follow a Boltzmann distribution. Scenarios which lead to a non-Boltzmann distribution include (i) strong shock waves, where the rapid temperature increase can preferentially excite certain vibrational modes; (ii) electrical discharges which can produce molecules in specific, highly excited vibrational states; and (iii) selective laser excitation, which can overpopulate a targeted energy level.

The validity of the assumption in these cases depends on how rapidly the energy distribution relaxes back to a Boltzmann shape. For N$_2$, this relaxation is governed by fast vibration-vibration (V-V) energy exchanges, which require only 10 to 100 collisions to establish a new Boltzmann distribution at a given $T_v$. This process is generally much faster than the vibrational-translational (V-T) energy exchange that equilibrates $T_v$ to the bulk gas temperature, $T$.

In plasma-assisted combustion, for instance, where gas temperatures are typically 500--3000~K, energy transfer to translation (V-T relaxation) is extremely slow, requiring $10^4$ to $10^6$ collisions. In contrast, V-V relaxation needs only 10--100 collisions---occurring in just 2--50 nanoseconds at 1~atm---making a Boltzmann distribution a valid assumption well before the second pulse starts. Because V-T relaxation times are typically much longer than the time spent between pulses, the vibrationally excited states created during one discharge remain for typically tens of pulses. Thus, most of the vibrational states at any given moment either during the discharges or not (except during the very first discharge) have had time to thermalize to a Boltzmann distribution. A similar dynamic governs laser-induced plasmas, where, despite an initially non-Boltzmann population from the laser pulse, rapid V-V thermalization re-establishes a Boltzmann distribution within a few nanoseconds. Consequently, the proposed model accurately approximates the heating of electrons from vibrationally excited states for these applications. This heating mechanism is significant here, potentially reaching 20--40\% of the reverse e-V cooling effect because the vibrational temperature reaches several thousand Kelvin (see Fig.~\ref{fig:contours_lossesgains}d).

In the post-shock region of hypersonic planetary entry flows, a similar argument holds. The V-V energy exchange mechanism is not strongly dependent on translational temperature, meaning a Boltzmann distribution is re-established within 10--100 collisions. For flight at Mach 10 to 30 and a dynamic pressure between 0.1 and 1~atm, this corresponds to a relaxation distance of only 0.1--5~mm. As this is generally much smaller than the shock-standoff distance, the assumption is valid for most of the post-shock flowfield.  Nonetheless, the assumption of a Boltzmann distribution can introduce modeling errors in certain regimes. For instance, in re-entry flight at low dynamic pressures near a small-radius leading edge, the relaxation time required to establish a Boltzmann distribution can approach or exceed the shock-standoff distance. In such cases, a state-to-state approach that tracks individual vibrationally excited species such as the one in \cite{jpca:2020:macdonald} may be more physically accurate.

However, this state-to-state approach presents significant drawbacks. First, the reaction rates governing energy exchange between these individual states are often not well known, introducing a source of physical modeling error that may be as large as that from the Boltzmann assumption itself. Second, the computational cost increases dramatically. Adding 20-40 species to account for the vibrationally excited states of N$_2$, NO, and other molecules would require solving a much larger system of transport equations. Implicit algorithms, necessary for such chemically reacting flows, have a computational cost that scales with the square or cube of the number of species.

This prohibitive cost often prevents the use of a sufficiently fine grid to reduce the numerical (or grid-induced) error to an acceptable level. Consequently, the numerical error associated with a coarse-grid state-to-state simulation could easily exceed the physical error from assuming a Boltzmann distribution on a fine grid. Since the goal is to minimize the sum of both numerical and physical errors, assuming a Boltzmann distribution is often the more accurate overall strategy in computationally demanding simulations, even in regimes where it introduces some physical inaccuracy.

\subsection{Validity of the Detailed Balance Principle}

While the principle of detailed balance is always valid for elementary reactions at the microscopic level due to microscopic reversibility, care must be taken when applying it to macroscopic rates. For instance, the use of the equilibrium constant through the detailed balance principle to determine reverse reaction rates can be inaccurate in systems under thermal non-equilibrium. This issue arises when the macroscopic forward and reverse reactions are governed by different temperatures. For example, in the associative ionization of nitrogen, $\text{N}+\text{N} \leftrightarrow \text{N}_2^+ + \text{e}^-$, the forward rate depends on the heavy particle (gas) temperature while the reverse rate depends on the electron temperature. In such cases, applying a single equilibrium constant to relate the two rates can lead to significant errors. To gauge this error, we can use the well-documented forward rate for this reaction by \cite{book:1990:park} and an equilibrium constant derived from the thermodynamic tables by \cite{nasa:2002:mcbride} to calculate the theoretical backward rate. When this calculated rate is compared to experimental data for the same reverse reaction by \cite{jgr:2004:sheehan}, the results differ by a factor of 2 to 5. This discrepancy effectively demonstrates the potential error when applying detailed balance to reactions where the forward and reverse paths are not governed by a single temperature. 

However, this limitation does not apply to the electron-impact vibrational excitation reactions under consideration herein, such as $\text{N}_2 + \text{e}^- \leftrightarrow \text{N}_2(v) + \text{e}^-$. For these reactions, both the forward (vibrational excitation) and backward (vibrational de-excitation) processes are a function of the electron temperature. Both the macroscopic forward and backward rates are the result of averaging the respective reversible microscopic rates over a distribution defined by the electron temperature. Thus, the equilibrium constant obtained by dividing the macroscopic forward rate by the macroscopic backward rate is only a function of one temperature: the electron temperature. Therefore, despite the system being in chemical and thermal non-equilibrium, applying the detailed balance principle to find the backward rate from the forward rate is applicable and does not lead to significant errors for the reactions involving vibrational excitation. 
}

\subsection{Comparison to Previous Electron Heating Models}

\begin{figure*}[!h]
     \centering
     \subfigure[\cite{jap:2010:shneider} model, Eq.~(\ref{eqn:scale_TediffTv})]{~~~~~~\includegraphics[width=0.405\textwidth]{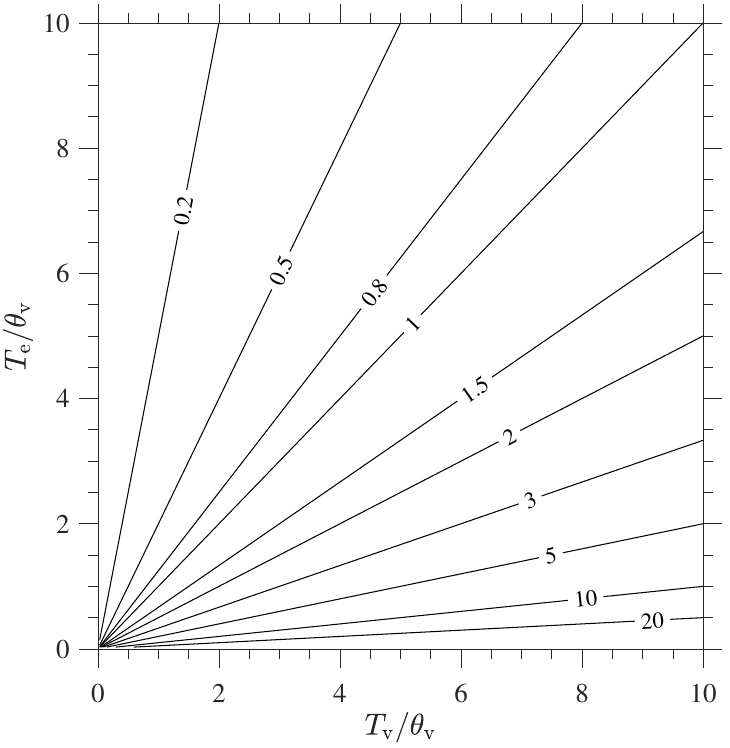}~~~~~~}
     \subfigure[\cite{jtht:2012:kim} model, Eq.~(\ref{eqn:scale_LandauTeller})]{~~~~~~\includegraphics[width=0.405\textwidth]{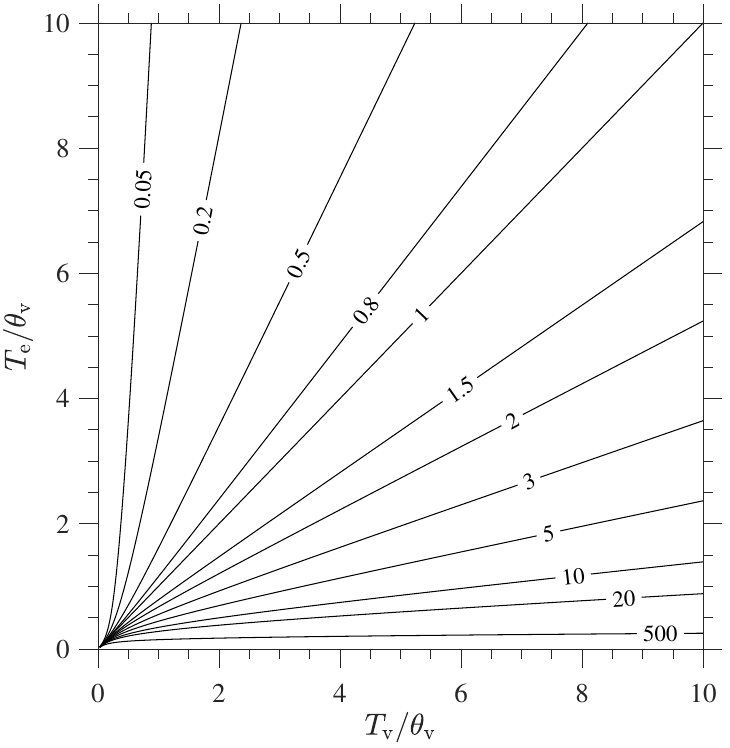}~~~~~~}
     \subfigure[\cite{jap:2019:peters} model, Eq.~(\ref{eqn:scale_Peters})]{~~~~~~\includegraphics[width=0.405\textwidth]{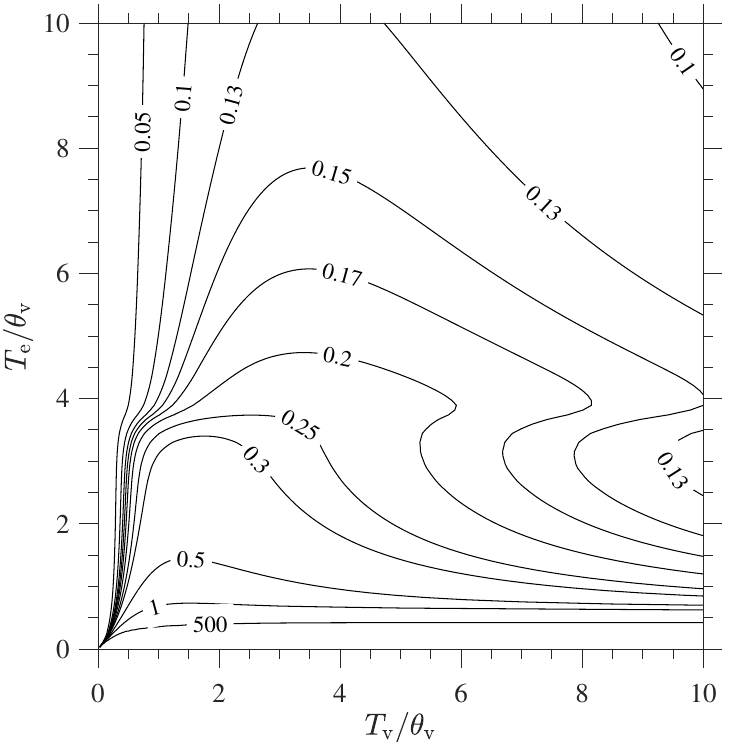}~~~~~~}
     \subfigure[Proposed model, Eq.~(\ref{eqn:scale_proposed})]{~~~~~~\includegraphics[width=0.405\textwidth]{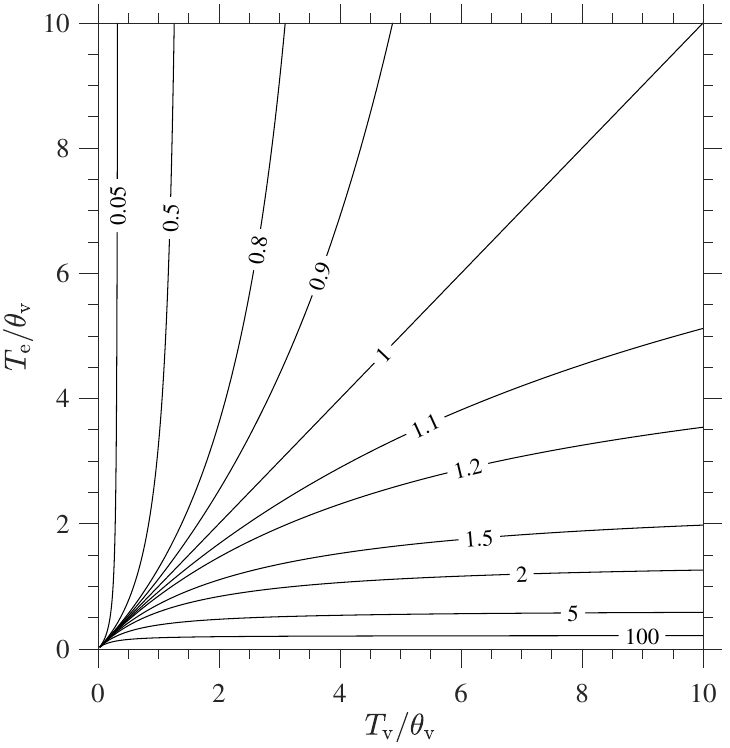}~~~~~~}
     \figurecaption{Ratio between V-e heating and e-V cooling for (a) the \cite{jap:2010:shneider} model Eq.~(\ref{eqn:scale_TediffTv}), (b) the \cite{jtht:2012:kim} model Eq.~(\ref{eqn:scale_LandauTeller}), (c) the \cite{jap:2019:peters} model Eq.~(\ref{eqn:scale_Peters}), and (d) the proposed model Eq.~(\ref{eqn:scale_proposed});  {(a), (b), and (d) apply to any molecule; (c) is limited to N$_2$.}}
     \label{fig:contours_lossesgains}
\end{figure*}

We outline briefly here the differences between our proposed electron heating model outlined in Eq.~(\ref{eqn:scale_proposed}) and previous inelastic electron heating models. One such prior model outlined in  \cite{jap:2010:shneider} assumed that the net electron heating-cooling rate scales with the difference between the vibrational and electron temperatures. This can be easily shown to lead to the following expression for the ratio of electron heating over electron cooling:
\begin{equation}
\frac{Q_{\rm v-e}}{Q_{\rm e-v}}
= \dfrac{T_{\rm v}}{T_{\rm e}} 
\label{eqn:scale_TediffTv}
\end{equation}
A second electron heating model that is thermodynamically consistent and that depends on the electron cooling rates is the one outlined in \cite{jtht:2013:farbar} and \cite{jtht:2012:kim}. Such  assumes a Landau-Teller form of the energy transfer rate between electrons and vibrational modes. In this case the ratio between inelastic electron heating rate and inelastic electron cooling rate becomes:
\begin{equation}
\displaystyle
\frac{Q_{\rm v-e}}{Q_{\rm e-v}} =  \dfrac{e_{\rm v}(T_{\rm v})}{e_{\rm v}(T_{\rm e})}
\label{eqn:scale_LandauTeller}
\end{equation}
A third electron heating model by \cite{jap:2019:peters} used detailed balancing to obtain the inelastic heating rates in standard form, considering the first 8 vibrational levels of molecular nitrogen. Similarly as done in this work, the fraction of ground state nitrogen in each vibrational level was computed as an harmonic oscillator obeying an equilibrium distribution. The latter leads to the following relation between heating and cooling rates:
%
\begin{equation}
\frac{Q_{\rm v-e}}{Q_{\rm e-v}} =  
\left[ 1 - \exp\left( - \frac{\theta_{\mathrm{v}}}{T_{\mathrm{v}}} \right) \right]
  \left(  \displaystyle\sum_{n=1}^{8} 
    \exp\left( - \frac{n \theta_{\mathrm{v}}}{T_{\mathrm{v}}} \right)
    k_n^{\mathrm{inv}} \mathcal{E}_n\right)
\left(
    \displaystyle\sum_{n=1}^{8} 
    k_n \mathcal{E}_n
\right)^{-1}
\label{eqn:scale_Peters}
\end{equation}
where the summation is made over the first eight vibrational level excitations from ground state and from detailed balance $k_{n}^{\rm inv}=k_{n}\exp\left(\mathcal{E}_n/k_{\rm B}T_{\rm e}\right)$. In Eq.~(\ref{eqn:scale_Peters}) the forward rates for vibrational excitation $k_n$ up to $n=8$ are obtained with BOLSIG+ using the Morgan cross-section data, extracted from the LXCat database by \cite{ppp:2017:pitchford}.

Despite being thermodynamically consistent, the first two previous ``phenomenological''  models by \cite{jap:2010:shneider} and by \cite{jtht:2012:kim} suffer from not satisfying the detailed balance principle. While the third model by \cite{jap:2019:peters} applies the correct detailed balance it is not guaranteed to ensure zero energy exchange when $T_{\rm e}=T_{\rm v}$.  In contrast, the model proposed herein shown in Eq.~(\ref{eqn:scale_proposed}) is not only thermodynamically consistent but also {approaches closely} the detailed balance principle because it is derived using the equilibrium constant. 

We present a comparison of the above three models with the proposed model in Fig.~\ref{fig:contours_lossesgains} on the basis of the ratio between inelastic heating and cooling. In hypersonic flows, the electron temperature is often less than the vibrational temperature. Thus, for such flows, the zone of interest is when  the heat-loss ratio is from 1 to 3. Comparing Fig.~\ref{fig:contours_lossesgains}a to Fig.~\ref{fig:contours_lossesgains}b, it can be seen that previous phenomenological models perform similarly in this zone of interest. In contrast, the proposed model leads to a much broader distance between contours indicating that it will lead to a larger difference between the electron and vibrational temperature when the former is lower than the latter. 

As well, as shown in Fig.~\ref{fig:contours_lossesgains}c, the \cite{jap:2019:peters} model fails to maintain a heat-loss ratio of 1 when the electron temperature is equal to the vibrational temperature. {As will be verified in the Test Cases section below, this is the reason why this model will not lead to $T_{\rm v}=T_{\rm e}$ at equilibrium.}

\subsection{Determination of Electron Swarm Parameters}

In the preceding sections we have employed two electron swarm parameters to obtain the inelastic electron energy gain-loss rates: the species reduced electric field  and the reduced electron mobility.  For all neutral species, such properties are obtained from the electron-temperature-dependent spline curve fits outlined in \cite{pf:2024:parent}. 

However, for the  N$_2$, NO, and O$_2$ molecules, we need to also incorporate the effect of the electron density on the reduced electric field. This is because the electron density influences the shape of the electron energy distribution function (EEDF) through electron-electron collisions and introduces density-dependent processes like stepwise ionization and Coulomb collisions. Such plays a minor role for very weakly-ionized plasmas where the ionization fraction is less than $10^{-5}$ or so. But this effect starts becoming very important at higher electron molar fractions encountered in hypersonic flows. 
\begin{figure}[!h]
     \centering
     \subfigure[N$_2$]{~~~\includegraphics[height=0.37\textwidth]{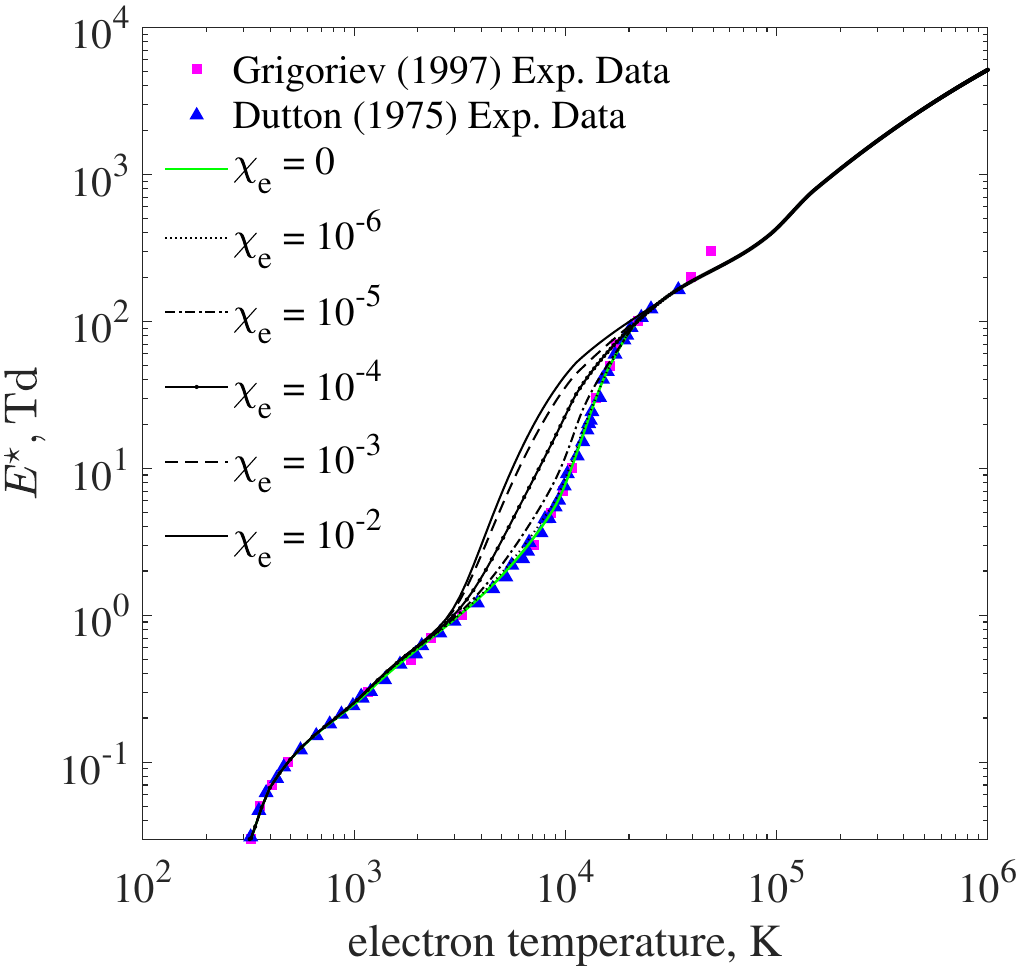}~~~}
     \subfigure[O$_2$]{~~~\includegraphics[height=0.37\textwidth]{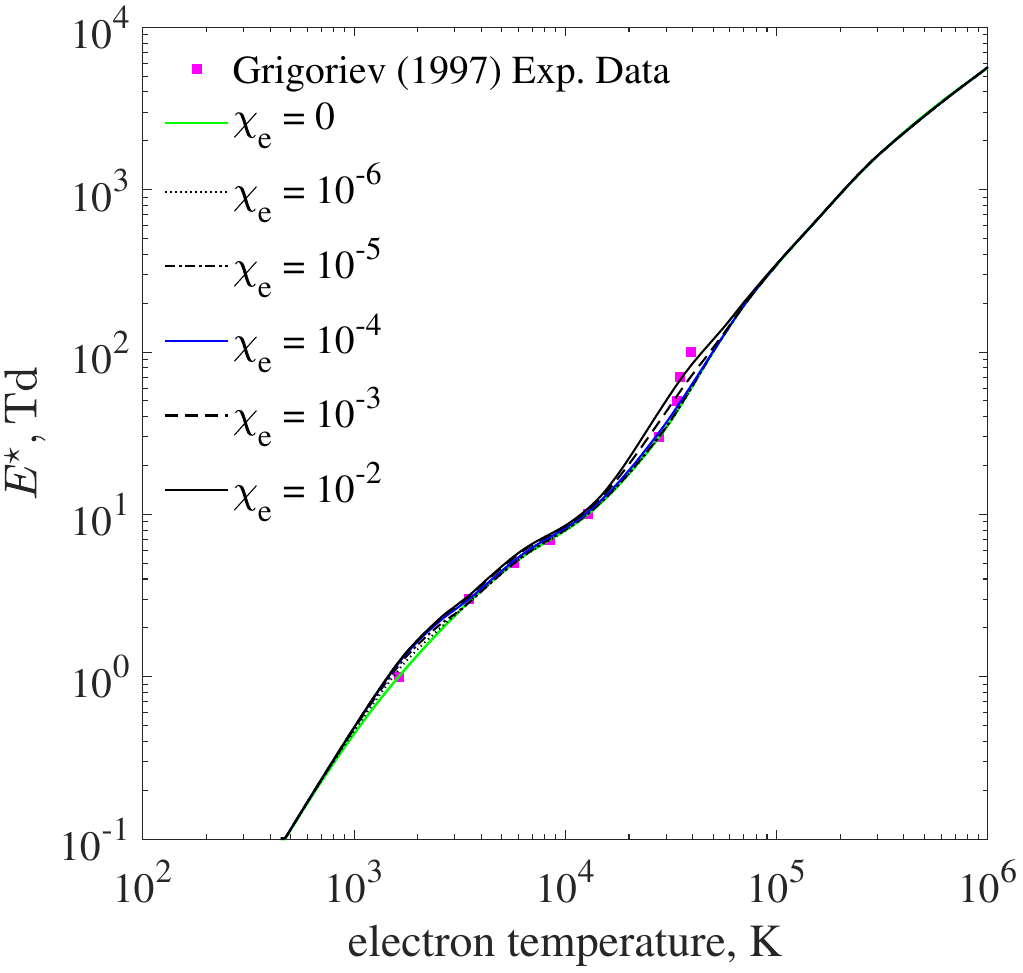}~~~}
     \subfigure[NO]{~~~\includegraphics[height=0.37\textwidth]{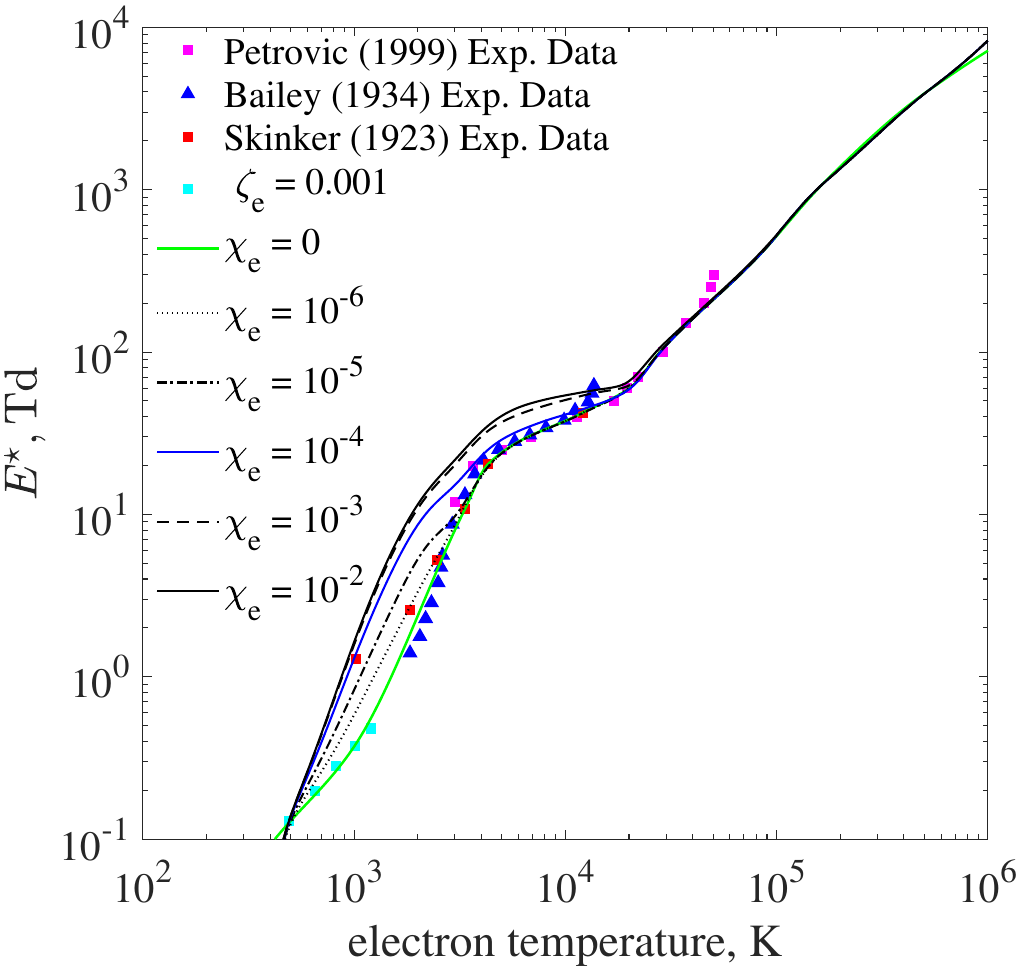}~~~}
     \figurecaption{Effect of electron molar fraction on the reduced electric field for (a) $\rm N_2$, (b) $\rm O_2$ and (c) $\rm NO$.}
     \label{fig:estar_chie}
\end{figure}
\begin{table*}[!ht]
  \center\fontsizetable
  \begin{threeparttable}
    \tablecaption{Spline control points giving the reduced electric field from the electron temperature and ionization degree\tnote{a}.}
    \label{tab:Estarfrom_Te_chie}
    \fontsizetable
 
    \begin{tabular*}{\textwidth}{@{}l@{\extracolsep{\fill}}cll@{}}
    
    \toprule
   {$\chi_{\rm e}$}~& Species &${\rm ln}~T_{\rm e}$ & ${\rm ln}~E^\star$  \\
        \midrule

   { 0   }  &  $\rm N_2$ &  \begin{minipage}[t]{0.4\textwidth}\raggedright  
 5.7836, 6.0067, 7.0566, 9.0580, 9.6956, 10.0010, 11.4313,
11.9302, 14.1582, 15.4249 \end{minipage}  & \begin{minipage}[t]{0.42\textwidth}\raggedright 
-51.8608, -51.0135, -49.5583, -46.7448, -44.4423, -43.7491,
-42.3992, -41.6848, -39.5410, -38.7385  \end{minipage}\\
~\\

    & $\rm O_2$ &  \begin{minipage}[t]{0.4\textwidth}\raggedright  
6.1549, 7.393, 8.6458, 9.4545, 10.2346, 10.9063, 11.1835, 11.8793, 12.655, 13.6691, 14.3029, 15.4249
 \end{minipage}  & \begin{minipage}[t]{0.42\textwidth}\raggedright 
-50.6569, -48.3543, -46.7448, -46.0517, -44.9531, -43.5718, -43.0406, -41.9779, -40.9153, -39.8524, -39.321, -38.7385
  \end{minipage}\\
~\\

     &  NO &  \begin{minipage}[t]{0.4\textwidth}\raggedright  
4.6052, 5.0387, 5.7384, 6.0438, 6.9078, 8.1163, 8.3684, 9.4004, 9.9245, 10.2589, 11.3016, 12.0194, 13.1593, 14.9141
 \end{minipage}  & \begin{minipage}[t]{0.42\textwidth}\raggedright 
-53.3605, -52.4569, -51.1678, -50.6572, -49.3446, -45.9734, -45.3367, -44.6079, -44.2446, -43.7196, -42.4083, -41.3594, -40.0476, -38.7385
\end{minipage}\\
~\\

  {  $10^{-6}$  } &  $\rm N_2$ &  \begin{minipage}[t]{0.4\textwidth}\raggedright  
5.7683, 5.9896, 6.5809, 7.1629, 8.0620, 8.4080, 9.0635, 9.8359, 10.2154, 11.3845, 12.2034, 13.8155 \end{minipage}  & \begin{minipage}[t]{0.42\textwidth}\raggedright 
-51.8608, -51.0557, -50.1105, -49.3864, -48.3393, -47.8592, -46.6400, -44.0774, -43.4726, -42.4529, -41.3656, -39.8080  \end{minipage}\\
~\\

    & $\rm O_2$ &  \begin{minipage}[t]{0.4\textwidth}\raggedright  
6.1522, 6.8503, 7.3536, 8.1003, 8.7526, 9.0764, 9.2107, 9.7428, 10.2573, 10.7492, 12.1589, 13.8155
\end{minipage}  & \begin{minipage}[t]{0.42\textwidth}\raggedright 
-50.6507, -49.2308, -48.3203, -47.3739, -46.6494, -46.3873, -46.278, -45.701, -44.9129, -43.8978, -41.5748, -39.7176
 \end{minipage}\\
~\\

     &  NO &  \begin{minipage}[t]{0.4\textwidth}\raggedright  
6.1522, 6.3101, 7.3536, 8.3067, 8.7526, 9.0764, 9.2107, 9.7428, 10.2573, 10.7492, 12.1589, 13.8155
 \end{minipage}  & \begin{minipage}[t]{0.42\textwidth}\raggedright 
-50.4801, -50.2342, -47.8325, -45.4518, -44.9853, -44.8002, -44.7282, -44.3947, -43.7214, -43.0928, -41.1794, -39.4755

\end{minipage}\\
~\\

  {  $10^{-5}$  } &  $\rm N_2$ &  \begin{minipage}[t]{0.4\textwidth}\raggedright  
5.7683, 5.9896, 6.5809, 7.1629, 8.0620, 8.4080, 9.0635, 9.6333, 10.2154, 11.3845, 12.2034, 13.8155 \end{minipage}  & \begin{minipage}[t]{0.42\textwidth}\raggedright 
-51.8608, -51.0557, -50.1085, -49.3801, -48.3084, -47.7657, -46.3350, -44.5139, -43.4726, -42.4529, -41.3656, -39.8080  \end{minipage}\\
~\\

    & $\rm O_2$ &  \begin{minipage}[t]{0.4\textwidth}\raggedright  
6.1522, 6.8503, 7.3536, 8.1003, 8.7526, 9.0764, 9.2107, 9.7428, 10.2573, 10.7492, 12.1589, 13.8155
\end{minipage}  & \begin{minipage}[t]{0.42\textwidth}\raggedright 
-50.6486, -49.2047, -48.2616, -47.3658, -46.6419, -46.374, -46.2694, -45.6947, -44.902, -43.8965, -41.5748, -39.7176
 \end{minipage}\\
~\\

     &  NO &  \begin{minipage}[t]{0.4\textwidth}\raggedright  
6.1522, 6.3101, 7.3536, 8.3067, 8.7526, 9.0764, 9.2107, 9.7428, 10.2573, 10.7492, 12.1589, 13.8155

 \end{minipage}  & \begin{minipage}[t]{0.42\textwidth}\raggedright 
-50.47, -50.2006, -47.3866, -45.4452, -44.9842, -44.8021, -44.7317, -44.4069, -43.7235, -43.0924, -41.1794, -39.4758
\end{minipage}\\
~\\

  {  $10^{-4}$  } &  $\rm N_2$ &  \begin{minipage}[t]{0.4\textwidth}\raggedright  
5.7683, 5.9896, 6.5809, 7.1629, 8.0620, 8.4080, 9.0635, 9.5975, 10.2154, 11.3845, 12.2034, 13.8155 \end{minipage}  & \begin{minipage}[t]{0.42\textwidth}\raggedright 
-51.8608, -51.0557, -50.1069, -49.3728, -48.2420, -47.4771, -45.6570, -44.3536, -43.4776, -42.4529, -41.3656, -39.8080 \end{minipage}\\
~\\

    & $\rm O_2$ &  \begin{minipage}[t]{0.4\textwidth}\raggedright  
 6.1522, 6.8503, 7.3536, 8.1003, 8.7526, 9.0764, 9.2107, 9.7428, 10.2573, 10.7492, 12.1589, 13.8155
 \end{minipage}  & \begin{minipage}[t]{0.42\textwidth}\raggedright 
-50.6478, -49.1915, -48.2261, -47.3118, -46.6086, -46.3478, -46.2461, -45.6619, -44.8429, -43.8833, -41.5748, -39.7176
 \end{minipage}\\
~\\

     &  NO &  \begin{minipage}[t]{0.4\textwidth}\raggedright  
6.1522, 6.3101, 7.3536, 8.3067, 8.7526, 9.0764, 9.2107, 9.7428, 10.2573, 10.7492, 12.1589, 13.8155

 \end{minipage}  & \begin{minipage}[t]{0.42\textwidth}\raggedright 
-50.4654, -50.1787, -46.837, -45.5396, -44.8367, -44.6856, -44.6356, -44.4024, -43.732, -43.0939, -41.1787, -39.4751
\end{minipage}\\
~\\

  {  $10^{-3}$  } &  $\rm N_2$ &  \begin{minipage}[t]{0.4\textwidth}\raggedright  
5.7683, 5.9896, 6.5809, 7.1629, 8.0620, 8.4080, 9.0635, 9.8359, 10.2154, 11.3845, 12.2034, 13.8155 \end{minipage}  & \begin{minipage}[t]{0.42\textwidth}\raggedright 
-51.8608, -51.0557, -50.1065, -49.3692, -48.1138, -47.0155, -45.0964, -43.9094, -43.4725, -42.4529, -41.3656, -39.8080 \end{minipage}\\
~\\

    & $\rm O_2$ &  \begin{minipage}[t]{0.4\textwidth}\raggedright  
6.1522, 6.8503, 7.3536, 8.1003, 8.7526, 9.0764, 9.2107, 9.7428, 10.2573, 10.7492, 12.1589, 13.8155
\end{minipage}  & \begin{minipage}[t]{0.42\textwidth}\raggedright 
-50.6477, -49.1885, -48.2132, -47.2757, -46.5669, -46.3193, -46.2194, -45.6058, -44.6918, -43.8101, -41.5748, -39.7176
 \end{minipage}\\
~\\

     &  NO &  \begin{minipage}[t]{0.42\textwidth}\raggedright  
6.1522, 6.3101, 7.3536, 8.3067, 8.7526, 9.0764, 9.2107, 9.7428, 10.2573, 10.7492, 12.1589, 13.8155

 \end{minipage}  & \begin{minipage}[t]{0.42\textwidth}\raggedright 
-50.464, -50.1705, -46.5977, -45.2394, -44.6054, -44.4747, -44.4278, -44.2838, -43.7226, -43.0961, -41.1794, -39.4765
\end{minipage}\\
~\\

  {  $10^{-2}$  } &  $\rm N_2$ &  \begin{minipage}[t]{0.4\textwidth}\raggedright  
5.7683, 5.9896, 6.5809, 7.1629, 8.0620, 8.4080, 9.0635, 9.8359, 10.2154, 11.3845, 12.2034, 13.8155 \end{minipage}  & \begin{minipage}[t]{0.42\textwidth}\raggedright 
-51.8608, -51.0557, -50.1065, -49.3686, -48.0548, -46.8277, -44.8902, -43.8230, -43.4632, -42.4529, -41.3656, -39.8080 \end{minipage}\\
~\\

    & $\rm O_2$ &  \begin{minipage}[t]{0.4\textwidth}\raggedright  
6.1522, 6.8503, 7.3536, 8.1003, 8.7526, 9.0764, 9.2107, 9.7428, 10.2573, 10.7492, 12.1589, 13.8155
 \end{minipage}  & \begin{minipage}[t]{0.42\textwidth}\raggedright 
-50.6477, -49.1881, -48.2113, -47.2633, -46.5492, -46.3089, -46.2091, -45.5539, -44.54, -43.6785, -41.5748, -39.7176
 \end{minipage}\\
~\\

     &  NO &  \begin{minipage}[t]{0.4\textwidth}\raggedright  
6.1522, 6.3101, 7.3536, 8.3067, 8.7526, 9.0764, 9.2107, 9.7428, 10.2573, 10.7492, 12.1589, 13.8155

 \end{minipage}  & \begin{minipage}[t]{0.42\textwidth}\raggedright 
-50.4638, -50.1693, -46.5479, -45.1622, -44.5042, -44.3959, -44.3626, -44.2493, -43.6633, -43.0712, -41.1793, -39.4747
\end{minipage}\alb

    \bottomrule
    \end{tabular*}
\begin{tablenotes}
\item[{a}] Notation and units: $T_{\rm e}$ is the electron temperature in Kelvin and $E^\star=|\vec{E}|/N$ is the reduced electric field in units of V$\rm ~m^2$. 
\end{tablenotes}
   \end{threeparttable}
\end{table*}

The method of incorporating the influence of electron-electron collisions on $E^\star$, shown in Fig.~\ref{fig:estar_chie} is as follows. First, we obtain the reduced electric field $E^\star_{\rm exp}$ using the experimental swarm data as a sole function of electron temperature. For $\rm N_2$ and $\rm O_2$ we use data from experiments outlined in \cite[Ch. 20]{book:1997:grigoriev} and from the comprehensive survey by \cite[Table 2.10]{jpcr:1975:dutton}. For NO, experimental data points are taken from  \cite[Fig.~2]{jop:1999:mechlinska},  \cite[Fig.\ 5]{jos:1934:bailey} and \cite[Fig.\ 1.32]{jos:1934:skinker}. Because there are indications that the BOLSIG+ results at such low electron temperature are not accurate, we here rather extrapolate from the experimental data to lower $T_{\rm e}$. The extrapolation of the $\rm O_2$ curve for $T_{\rm e}<1000$~K is such that there is good agreement with experimental data of air at low electron temperature (as will be shown later in the first test case). {The extrapolation of the NO curve for $T_{\rm e}<1000$~K is such that the reduced electric field is computed from the following relation function of the electron energy loss function $\zeta_{\rm e} $ (the mean fraction of electron kinetic energy that the electron loses in one collision), as outlined in \cite{pf:2024:parent}:
\begin{equation}
E^\star = \sqrt{\frac{3  k_{\rm B} T_{\rm e}\zeta_{\rm e} }{2 m_{\rm e} (\mu_{\rm e}^\star)^2}}     
\label{eqn:Estar_from_zetae}
\end{equation}
At a temperature of less than 1000~K, we here set $\zeta_{\rm e}$ to 0.001 because electrons in a molecular gas at low temperature mostly dissipate their energy by exciting the vibrational and rotational energy modes and such leads to a $\zeta_{\rm e}$ varying between $10^{-3}$ and $10^{-2}$ as outlined in \cite[page 17]{book:1991:raizer}. Such is one or two orders of magnitude higher than the electron energy loss by elastic collisions ($\zeta_{\rm e}=2 m_{\rm e}/m_{\rm n} \approx 4\times10^{-5}$).}

{We note that the above $E^\star$ extrapolation for $T_{\rm e}<1000$~K introduces some error because the value of $\zeta_{\rm e}$ is prescribed rather than obtained from swarm experiments. However, for the two hypersonic reentry test cases that will be shown in this text, the electron temperature remains above this threshold, so that the extrapolated portion of the curve is not sampled. Thus, for the cases discussed here we do not expect additional errors arising from this extrapolation, but acknowledge the uncertainties in sampling data from the splines below 1000 K.}

Next, the reduced electric field is computed for ionization degrees $\chi_{\rm e}\equiv N_{\rm e}/N$ ranging from zero to $10^{-2}$ numerically using BOLSIG+. The cross-section data required for BOLSIG+ is sourced from the Phelps database for the NO species and from the Morgan database for $\rm N_2$ and $\rm O_2$. Both the Morgan and Phelps cross-sectional data are extracted from the LXCat database by \cite{ppp:2017:pitchford}. We then start from the spline fits to swarm experimental data of the reduced electric field to add the difference in $E^\star$ taking the numerical solution with $\chi_{\rm e}=0$ as a reference. Note that because the electron energy loss rates in Eq.~(\ref{eqn:Qinelastic_cooling}) scale as $(E^\star)^2$, a suitable addition process to obtain $E^{\star}$ as a function of $T_{\rm e}$ and $\chi_{\rm e}$ is as follows:
\begin{equation}
\displaystyle
E^\star(T_{\rm e},\chi_{\rm e}) = \left[\left(E^\star(T_{\rm e})\right)^2_{\rm exp} + \left(E^\star(T_{\rm e},\chi_{\rm e})\right)^2_{\rm bol}  - \left(E^\star(T_{\rm e},\chi_{\rm e}=0)\right)^2_{\rm bol} \right]^\frac{1}{2}
\label{eqn:Estar_from_chie}
\end{equation}
Given the latter for a specified number of ionization fraction values (see Fig.\ \ref{fig:estar_chie}), it is straightforward to interpolate at any given $\chi_{\rm e}$ found at each point in the computational domain during the convergence process. In Table \ref{tab:Estarfrom_Te_chie} we list the required spline control points to evaluate $E^\star(T_{\rm e},\chi_{\rm e})$ for the species $\rm N_2$, $\rm O_2$ and NO.

\begin{figure}[!t]
    \centering
     \includegraphics[width=0.38\textwidth]{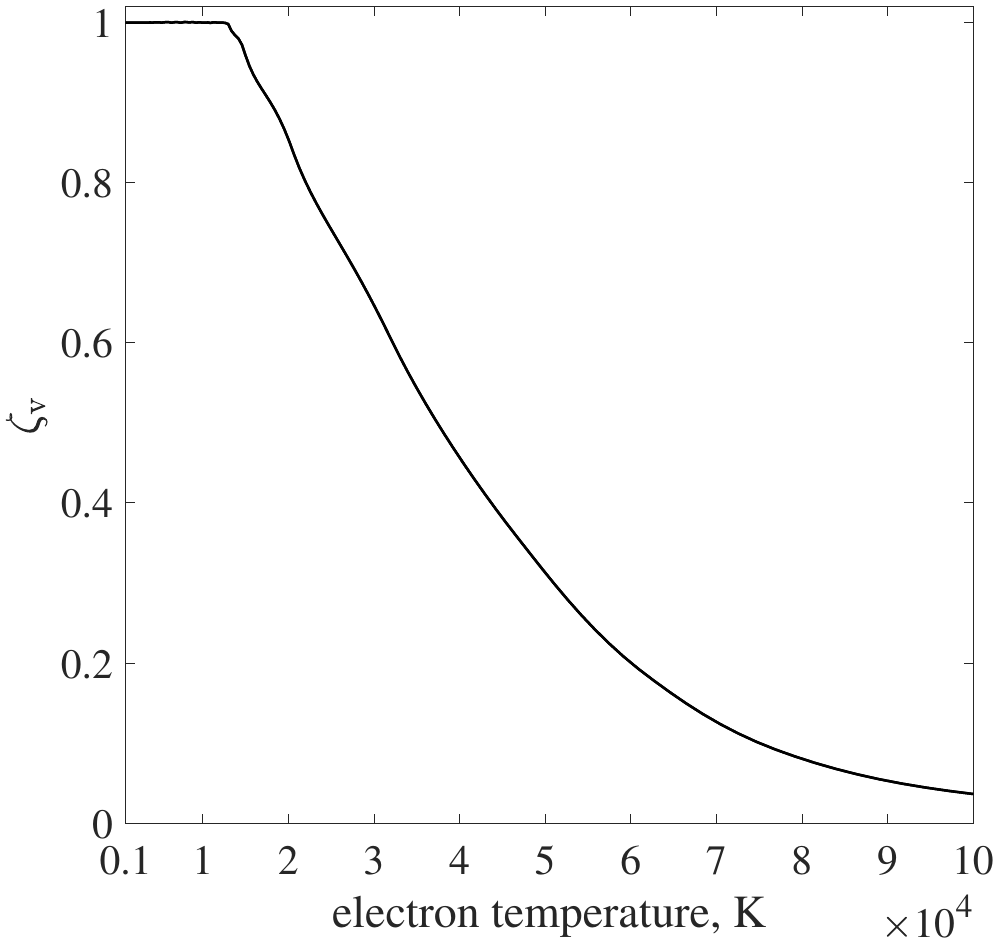}
     \figurecaption{Fraction of electron energy consumed in the excitation
of the $\rm N_2$ vibrational energy.}
     \label{fig:energyloss_N2}
\end{figure}

\subsection{Model Implementation for Molecular Nitrogen}

\begin{table*}[!ht]
\fontsizetable
\begin{center}
\begin{threeparttable}
\tablecaption{Spline control points giving the fraction of the electron energy lost in the excitation of the N$_2$ vibrational energy.}
    \begin{tabular*}{0.99\textwidth}{@{}l@{\extracolsep{\fill}}l@{}}
    \toprule
    $T_{\rm e}$, K & $\zeta_{\rm v}$  \\
        \midrule
   \begin{minipage}[t]{0.52\textwidth}\raggedright  
300,	10405,	12758,	21673,	38350,	55346,	83199,	115277,	188009,	407883,	638657,	1000000 \end{minipage}  & \begin{minipage}[t]{0.48\textwidth}\raggedright 
1.0000,	0.99970,	0.99809,	0.80839,	0.483534,	0.24692,	0.07006,	0.02239,	0.005022,	0.000793,	0.0003375,	0.000158

\end{minipage}\alb
    \bottomrule
    \end{tabular*}
\label{tab:zetav_table_spline}
\end{threeparttable}
\end{center}
\end{table*}

{Because the sole source of electron heating is here assumed to be due to the vibrationally excited N$_2$ molecules losing energy to reach the ground state, we should thus limit the application of the proposed model to the N$_2$ vibrationally excited states. We can easily find the energy transfer from the electrons to the vibrational modes of nitrogen noting that this corresponds to the product between the fraction of electron energy loss consumed in the excitation of the nitrogen vibrational energy (denoted herein as $\zeta_{\rm v}$) and the total electron cooling due to all inelastic processes outlined in Eq.~(\ref{eqn:Qinelastic_cooling}):
\begin{equation}
Q_{\rm e-v} =   \zeta_{\rm v} |C_{\rm e}| N_{\rm e} N_{\rm N_2} (\mu_{\rm e}^\star)_{\rm N_2} \left( (E^\star_{\rm N_2})^2 -  \frac{3  k_{\rm B}    (T_{\rm e}-T_{\rm ref})}{ m_{\rm N_2} (\mu_{\rm e}^\star)^2_{\rm N_2}} \right) 
\end{equation}
where $\theta_{\rm v}$ is the nitrogen characteristic vibration temperature which is set to 3353~K, as suggested in \cite{book:1962:barrow}. We can then find the electron heating by multiplying the latter by the ratio between the heating and cooling processes outlined in  Eq.~(\ref{eqn:scale_proposed}):
\begin{equation}
Q_{\rm v-e} =   \zeta_{\rm v} |C_{\rm e}| N_{\rm e} N_{\rm N_2} (\mu_{\rm e}^\star)_{\rm N_2} \left( (E^\star_{\rm N_2})^2 -  \frac{3  k_{\rm B}    (T_{\rm e}-T_{\rm ref})}{ m_{\rm N_2} (\mu_{\rm e}^\star)^2_{\rm N_2}} \right)  \exp\left(\frac{\theta_{\rm v}}{T_{\rm e}}-\frac{\theta_{\rm v}}{T_{\rm v}} \right)
\label{eqn:Qve_implementation}
\end{equation}
}We here determine $\zeta_{\rm v}$ as the ratio of energy losses due to all vibrational excitation processes of $\rm N_2$ and the energy losses due to all $\rm N_2$ electron-impact processes (vibrational excitation, electronic excitation, dissociative excitation and ionization) but excluding rotational excitation. We do not include rotational excitation cross-sectional data due to the large uncertainties in the data associated with this process. The latter reaction rates are obtained by integration of the EEDF using BOLSIG+ with the Morgan set of $\rm N_2$ cross sections sourced from the \cite{cpp:2012:pancheshnyi} LXCat collaborative database. The latter fraction of electron energy lost in exciting the nitrogen vibrational energy is shown in Fig.~\ref{fig:energyloss_N2} with corresponding spline-fitted data given in Table~\ref{tab:zetav_table_spline}.


{
\subsection{Extension to Multiple Vibrationally-Excited Species}

While the model derivation and validation have centered on molecular nitrogen (N$_2$), the proposed methodology is generalizable and can be extended to incorporate the effects of other vibrationally-excited species. This extension involves a two-step process.

First, to account for the vibrational non-equilibrium of an additional species, such as nitric oxide (NO), a corresponding vibrational energy transport equation must be added to the governing system of equations. This new equation, which would be analogous to Eq.~\eqref{eqn:vibrationalenergytransport} for N$_2$, is necessary to solve for the vibrational temperature associated specifically with that species (e.g., $T_{\rm v}^\text{NO}$). This results in a multi-vibrational temperature model where the system tracks $T_{\rm v}^{\text{N}_2}$, $T_{\rm v}^{\text{NO}}$, and so on.

Second, the electron energy transport equation, Eq.~\eqref{eqn:electronenergytransport}, must be updated to include the additional electron heating term associated with this newly-tracked species. The total vibrational-electron heating term, $Q_{\rm v-e}$, would then be a summation of the contributions from each molecule. For a system including N$_2$ and NO, the heating term would be $Q_{\rm v-e}^{\text{N}_2} + Q_{\rm v-e}^{\text{NO}}$. The heating from NO, $Q_{\rm v-e}^{\text{NO}}$, would be calculated using the same formulation presented in Eq.~\eqref{eqn:Qve_implementation}, but adapted for the NO molecule. This adaptation requires two key parameters for NO: its characteristic vibrational temperature, $\theta_{\rm v}$, which is 2719~K, and the fraction of electron energy consumed in the excitation of its vibrational energy, $\zeta_{\rm v}$. This fraction must be determined using BOLSIG+ with the appropriate cross-section data for NO, following the same procedure used for N$_2$.

By following these steps, the thermodynamically consistent heating model can be systematically and flexibly applied to more complex, multi-species plasma environments where more than one molecule contributes significantly to vibrational-electron energy exchange.}

\section{Numerical Methods}

All results presented herein were obtained using the in-house CFDWARP (Computational Fluid Dynamics, WAves, Reactions, Plasmas) code. CFDWARP has the capability to integrate efficiently with aerodynamic-size time steps the physical system outlined in the previous sections.

The coupled system of drift-diffusion, fluid transport, and Gauss's law equations is notoriously stiff, posing a significant numerical challenge due to the wide range of physical timescales involved. To effectively manage this stiffness and ensure efficient convergence to a steady state, a multi-faceted numerical strategy is employed. The governing equations are first recast to a different form following the approach outlined in \cite{book:2022:parent} and \cite{jcp:2015:parent}. Such permits much higher integration time steps to be used by improving the mathematical conditioning of the system without altering the underlying physics.
The time-integration to a steady-state solution is achieved with tailored algorithms for different parts of the system. The discretized fluid flow transport equations are converged using a computationally efficient Diagonally Dominant Alternate Direction Implicit (DDADI) block-implicit method by \cite{aiaaconf:1987:bardina} chosen because of its good high-frequency damping characteristics. For the elliptic Poisson equation for the electric potential, which must be solved at each step, a combination of a Successive Over-Relaxation (SOR) algorithm \cite{gen:douglas} and an Iterative Modified Approximate Factorization (IMAF) algorithm {by} \cite{cf:2001:maccormack} is used to accelerate convergence.

For the spatial discretization of the governing equations, the convective fluxes are approximated using the \cite{jcp:1981:roe} scheme, an upwind method chosen for its shock-capturing capabilities in hypersonic flows. To extend this scheme to second-order accuracy in smooth flow regions while simultaneously preventing spurious, non-physical oscillations near discontinuities, the \cite{cfd:vanleer} Total Variation Diminishing (TVD) flux limiter is utilized. A known potential issue with Roe's scheme at high Mach numbers is the formation of numerical artifacts known as carbuncles. Such is suppressed by applying an eigenvalue conditioning technique based on the local Peclet number, as outlined in \cite{aiaa:2017:parent}. This ensures robustness without compromising the sharp resolution of the viscous layer. The diffusion terms are discretized using a standard and stable second-order accurate central-differencing stencil.

\begin{figure}[!t]
     \centering
     \includegraphics[width=0.38\textwidth]{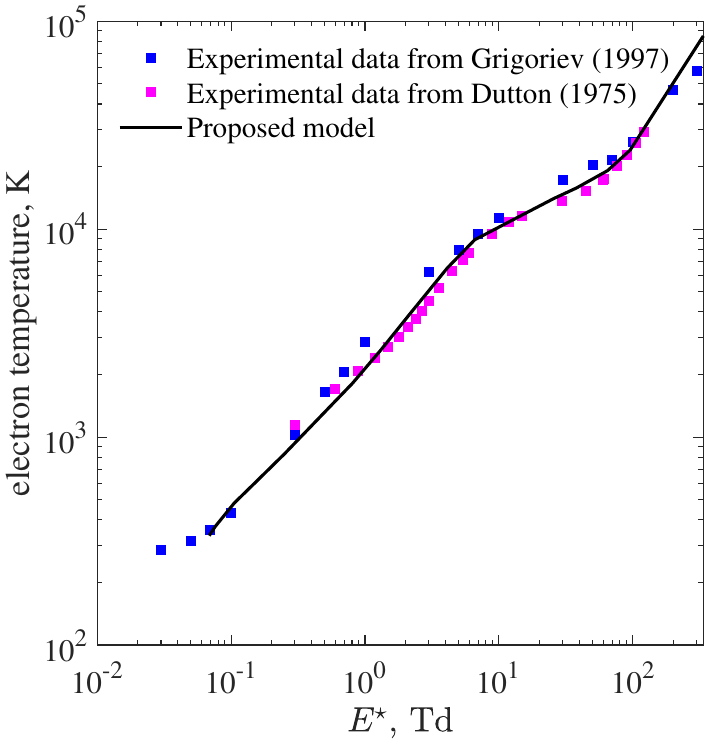}
     \figurecaption{Electron temperature as a function of the reduced electric field for the uniformly applied electric field test case.}
     \label{fig:appliedE_Te_vs_Estar}
\end{figure}

\section{Test Cases}

We now test the accuracy of our proposed electron energy model. {Four test cases are presented: the uniform electric field applied to a uniform plasma test case, the RAM-C-II and OREX reentry flight test cases, and a thermal equilibrium test case.} For the re-entry test cases we test the validity of the proposed model against  the \cite{jap:2019:peters} model and the \cite{jtht:2012:kim} model. For the thermal equilibrium plasma test case, we include an additional comparison to the phenomenological model of \cite{jap:2010:shneider}. 

\subsection{Uniform Electric Field Test Case}

The objective of this first test case is to verify the accuracy of the proposed electron energy equation at predicting the electron temperature of a weakly-ionized plasma subject to an externally applied electric field. To do so, we will apply a uniform electric field on a plasma flow entering the computational domain with a speed of 2~km/s, a temperature of 300 K, and a pressure of 1~atm. The molar fractions of electrons and $\rm N_2^+$ are set to $10^{-10}$. We apply supersonic inflow and outflow boundary conditions on the left and right edges of the domain. Symmetry boundaries are specified on the top and bottom edges of the 2D domain 3~cm in length and 1~cm in height. Next, we apply a uniform power deposition $\vec{E}\cdot \vec{J}_{\rm e}$ within the range $10^2$ to $10^9$ W/m$^3$. Following the derivation in \cite{pf:2024:parent}, the reduced electric field in the uniform plasma can be shown to correspond to:
\begin{equation}
E^\star=\frac{1}{N}\sqrt{\frac{\vec{E}\cdot\vec{J}_{\rm e}}{\mu_{\rm e} N_{\rm e} |C_{\rm e}|}} 
\label{eqn:EstarfromEdotJe}
\end{equation}

We employ here the \cite{nasa:1990:gupta} transport coefficients but with the reduced electron mobility for $\rm N_2$ and $\rm O_2$ taken from \cite{pf:2024:parent}, obtained from electron swarm experiments. This substitution is made because the Gupta–Yos model is not valid at very high electron temperatures exceeding 30,000~K.

From the converged solution at each specified power deposition, we measure the electron mobility, number densities and electron temperature at the outflow plane. Then, we compute the reduced electric field $E^\star$ as given in Eq.~(\ref{eqn:EstarfromEdotJe}). The results are compared with experimental data of $\rm N_2$ and $\rm O_2$ from \cite[Ch. 20]{book:1997:grigoriev} (mixture of 79\% nitrogen and 21\% oxygen) and experimental dry air swarm data from \cite{jpcr:1975:dutton} with no presence of carbon dioxide. This is shown in Fig.\ \ref{fig:appliedE_Te_vs_Estar} with a close match to experimental air data and an error not exceeding 25\%.  The good agreement found in this test case is a result of all relevant inelastic energy loss processes (vibrational excitation, electronic excitation, Townsend ionization, etc) being included within the electron energy losses source terms.

\begin{figure}[t]
     \centering
     \includegraphics[width=0.38\textwidth]{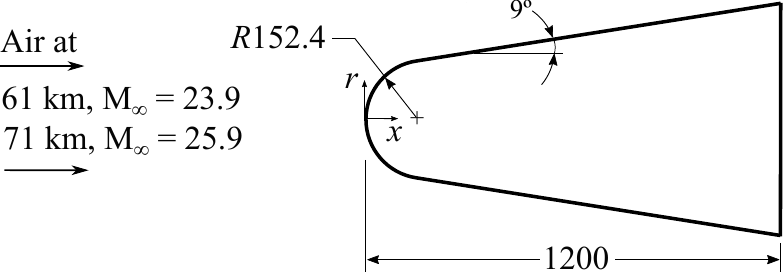}
     \figurecaption{Problem setup for the RAM-C-II test cases; dimensions in mm.}
     \label{fig:setup_RAMCII}
\end{figure}

\begin{figure}[!h]
     \centering
     \subfigure[]{~~~\includegraphics[width=0.30\textwidth]{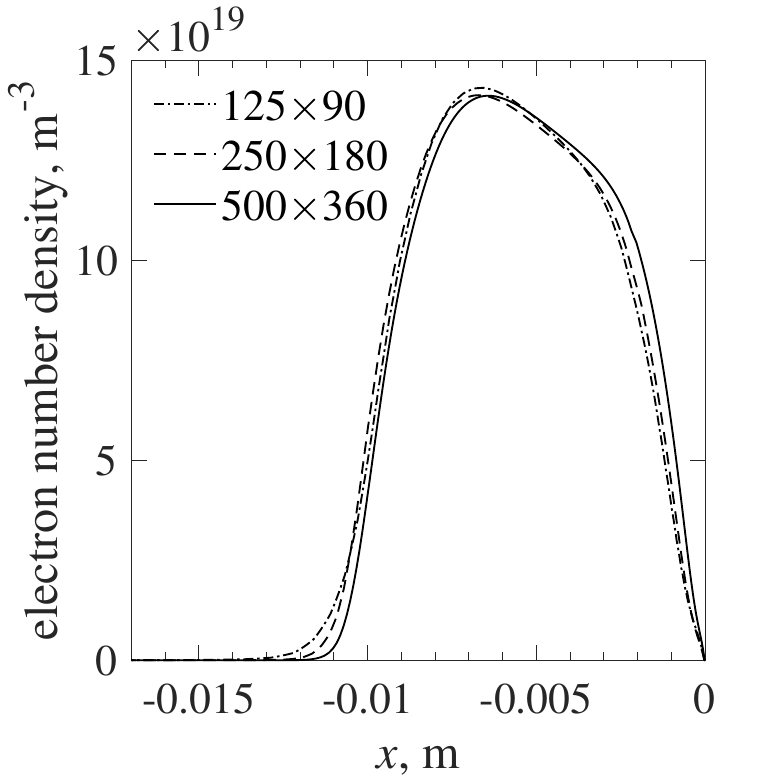}~~~}
     \subfigure[]{~~~\includegraphics[width=0.30\textwidth]{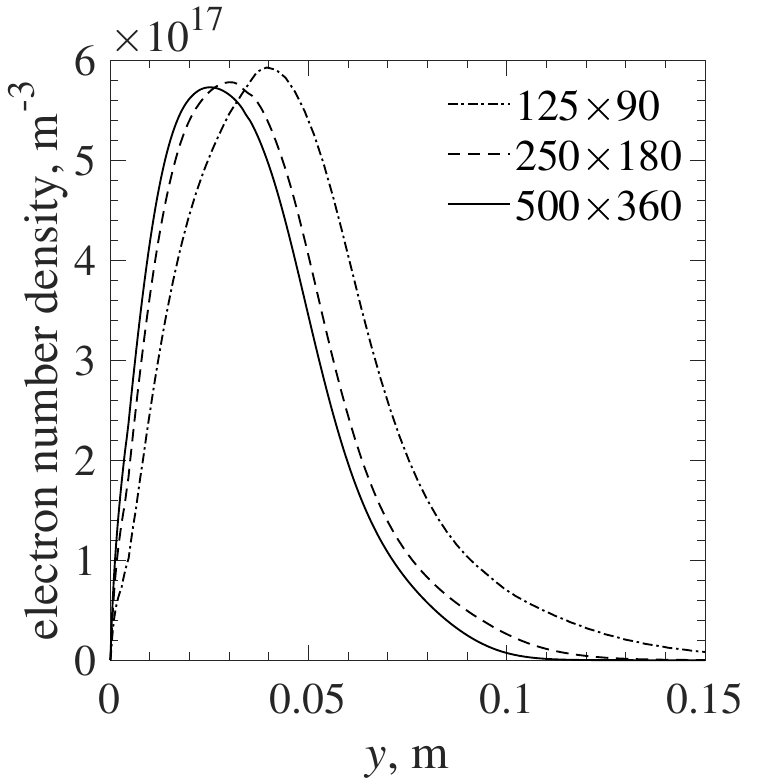}~~~}
     \figurecaption{ Effect of grid size on (a) electron number density along the stagnation streamline and (b) electron number density 0.6 meters aft from the
leading edge, for the 61 km altitude case.}
     \label{fig:RAMCII_grid_convergence_Ne}
\end{figure}

\begin{figure*}[!ht]
     \centering
     \subfigure[61 km]{\includegraphics[width=0.46\textwidth]{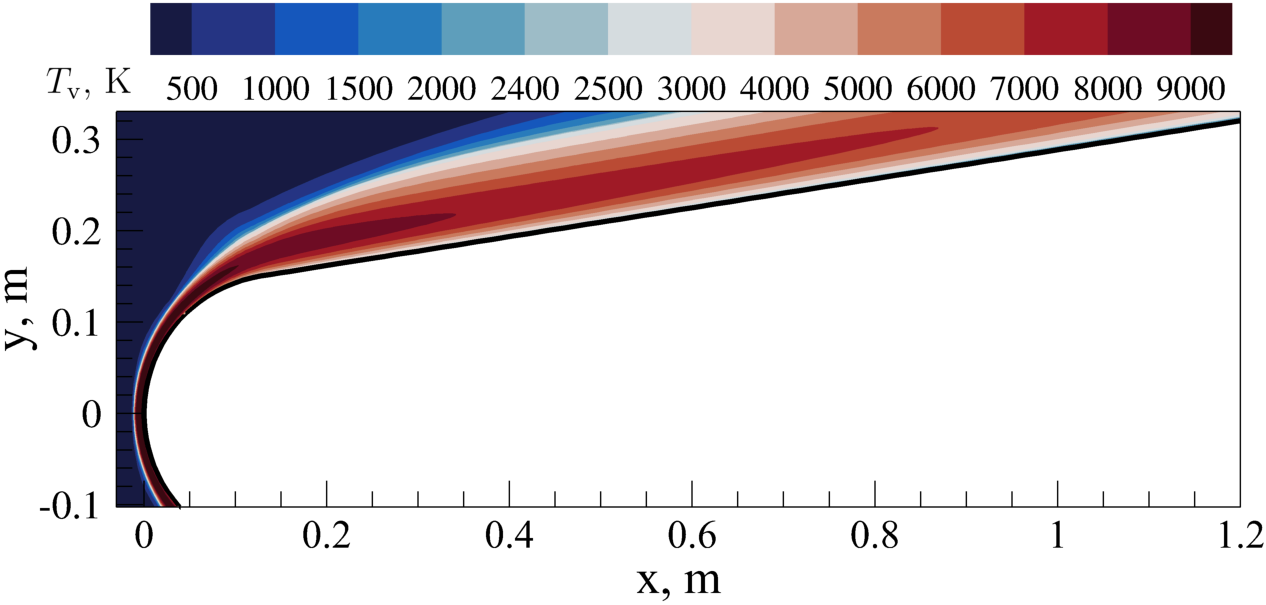}~~~}
     \subfigure[71 km]{\includegraphics[width=0.46\textwidth]{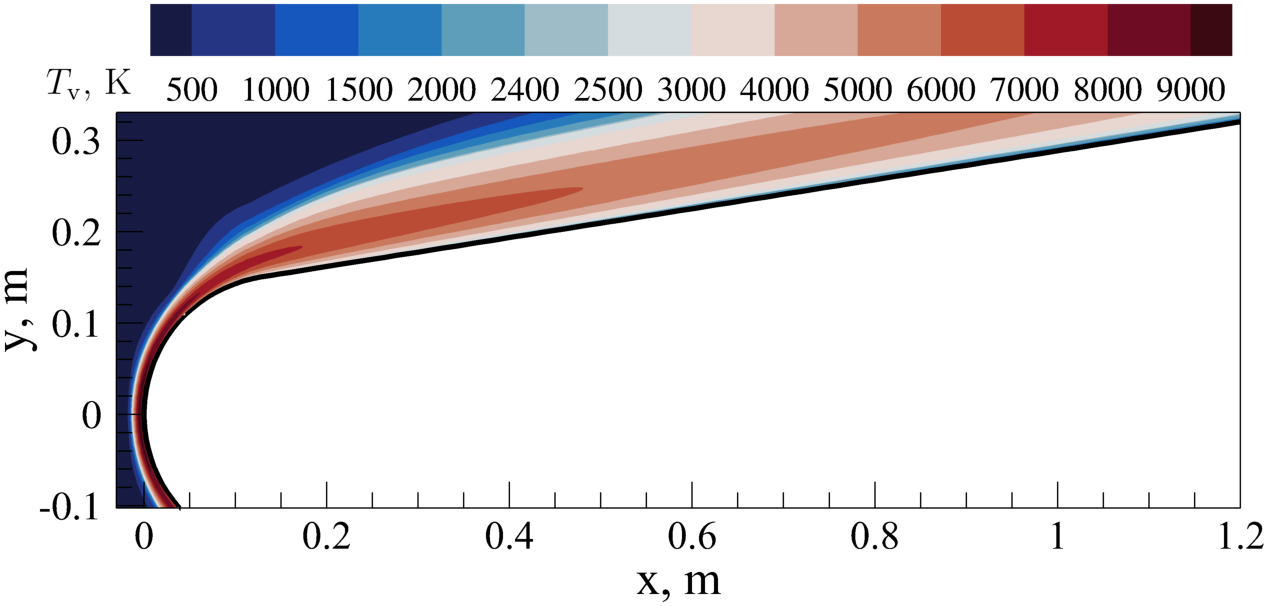}~~~}
     \figurecaption{Vibrational temperature around the RAM-C-II body for a) 61 km and b) 71 km altitude cases.}
     \label{fig:RAMCII_Tv_comparison}
\end{figure*}

\begin{figure*}[ht]
     \centering
     \subfigure[61 km, Kim et al. (2012) model]{~~~\includegraphics[width=0.46\textwidth]{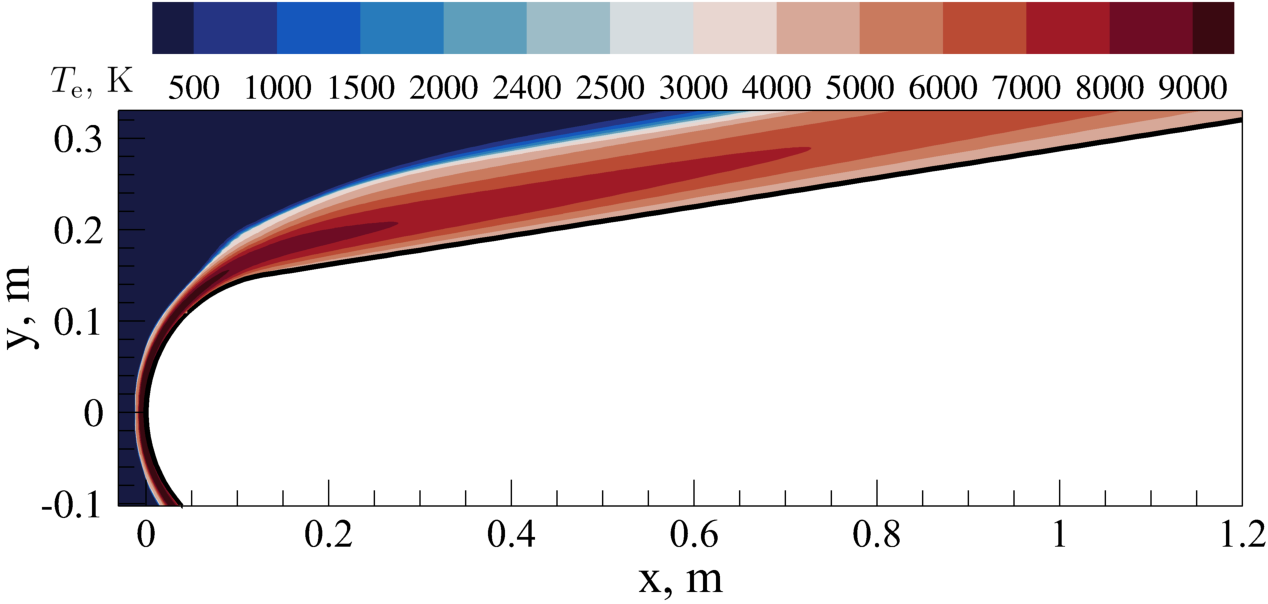}}
     \subfigure[71 km, Kim et al. (2012) model]{~~~\includegraphics[width=0.46\textwidth]{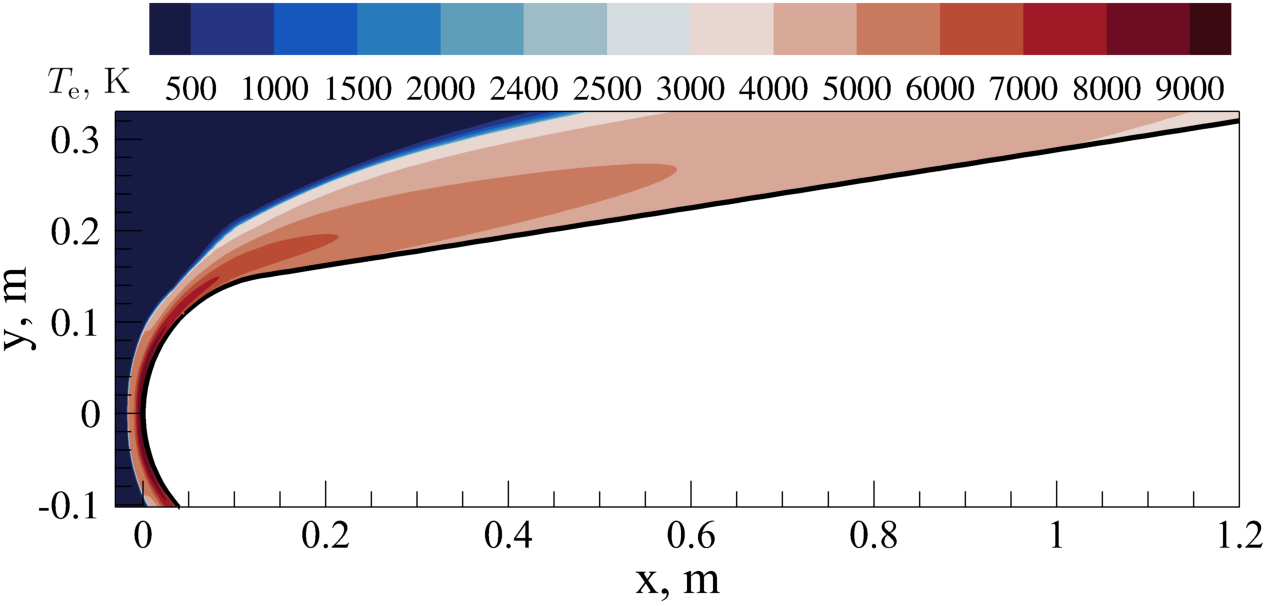}}
     \subfigure[61 km, Peters et al. (2019) model]{~~~\includegraphics[width=0.46\textwidth]{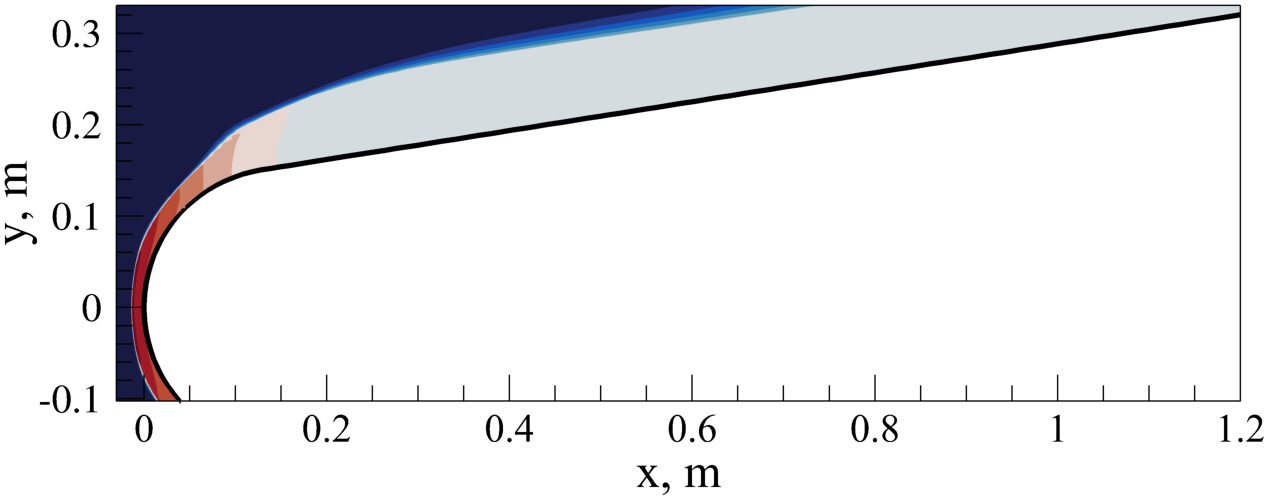}}
     \subfigure[71 km, Peters et al. (2019) model]{~~~\includegraphics[width=0.46\textwidth]{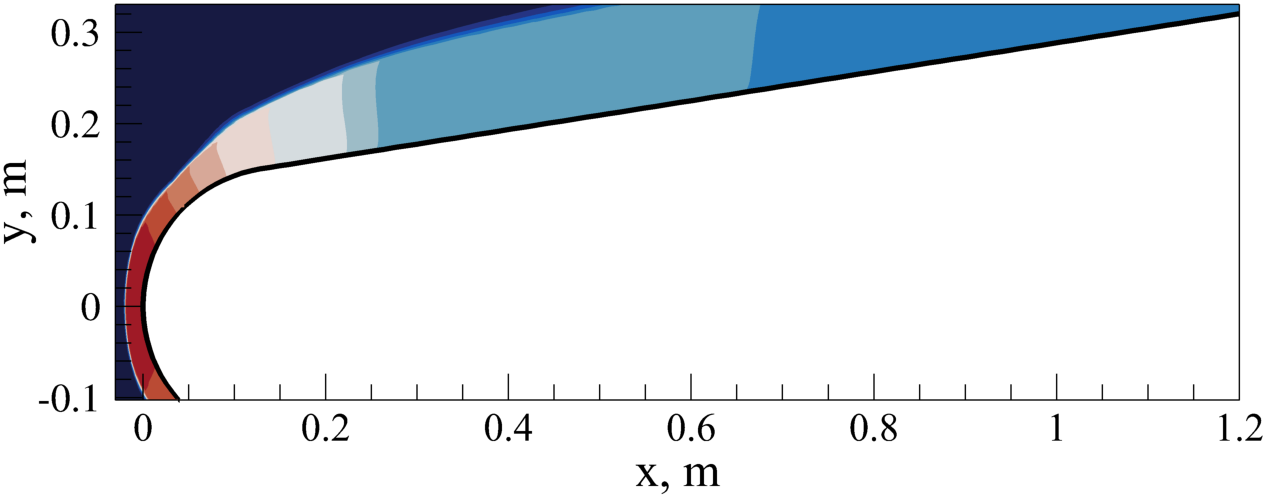}}
     \subfigure[61 km, Proposed model]{~~~\includegraphics[width=0.46\textwidth]{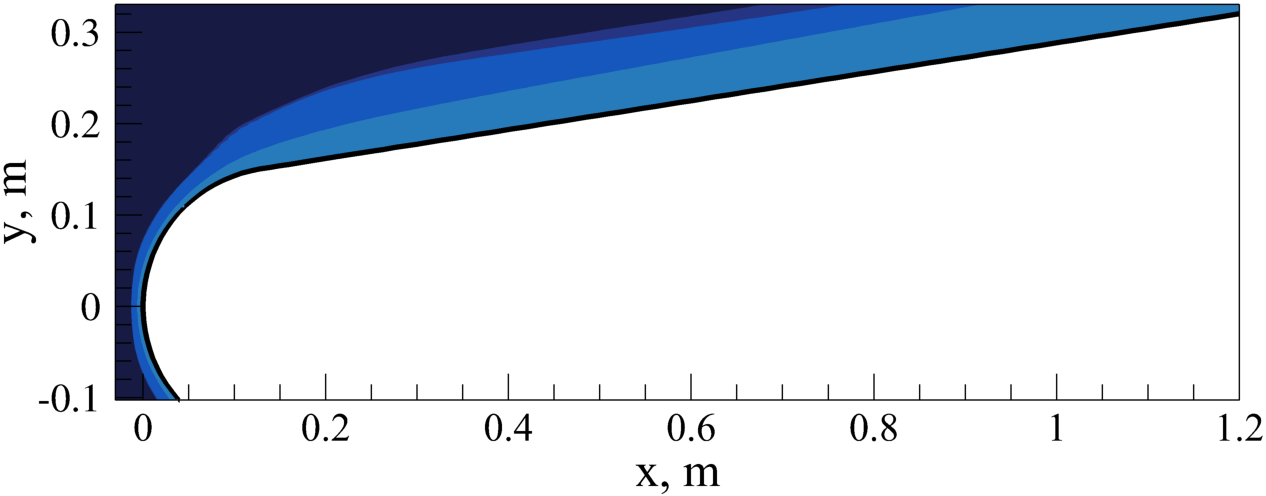}~~~}
     \subfigure[71 km, Proposed model]{~~~\includegraphics[width=0.46\textwidth]{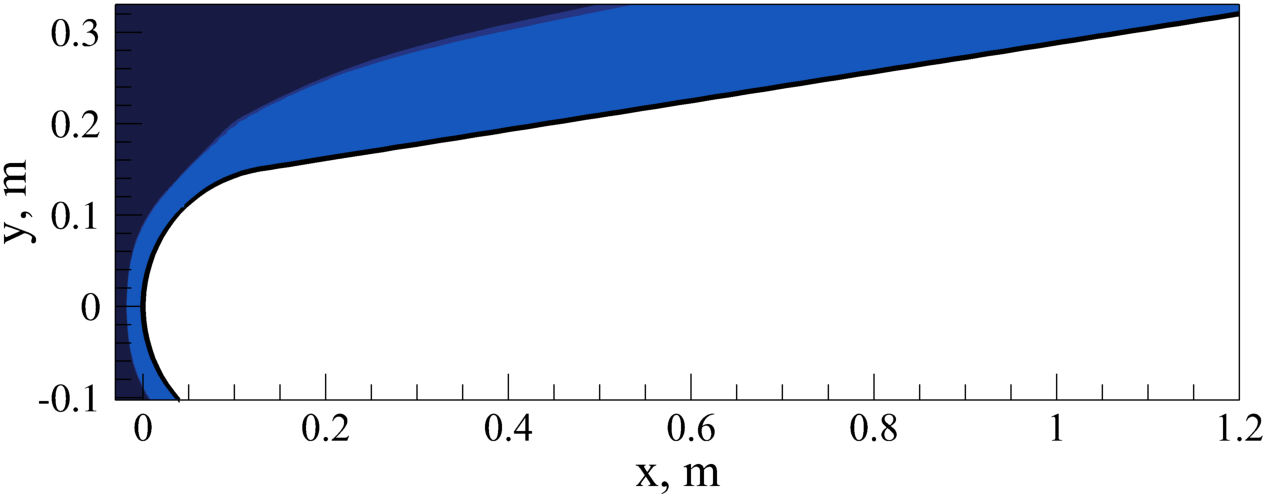}}
     \figurecaption{Electron temperature around the RAM-C-II body for the 61 km and 71 km altitude cases obtained with different electron heating models.}
     \label{fig:RAMCII_Te_comparison}
\end{figure*}

\begin{figure}[!h]
     \centering
     \includegraphics[width=0.40\textwidth]{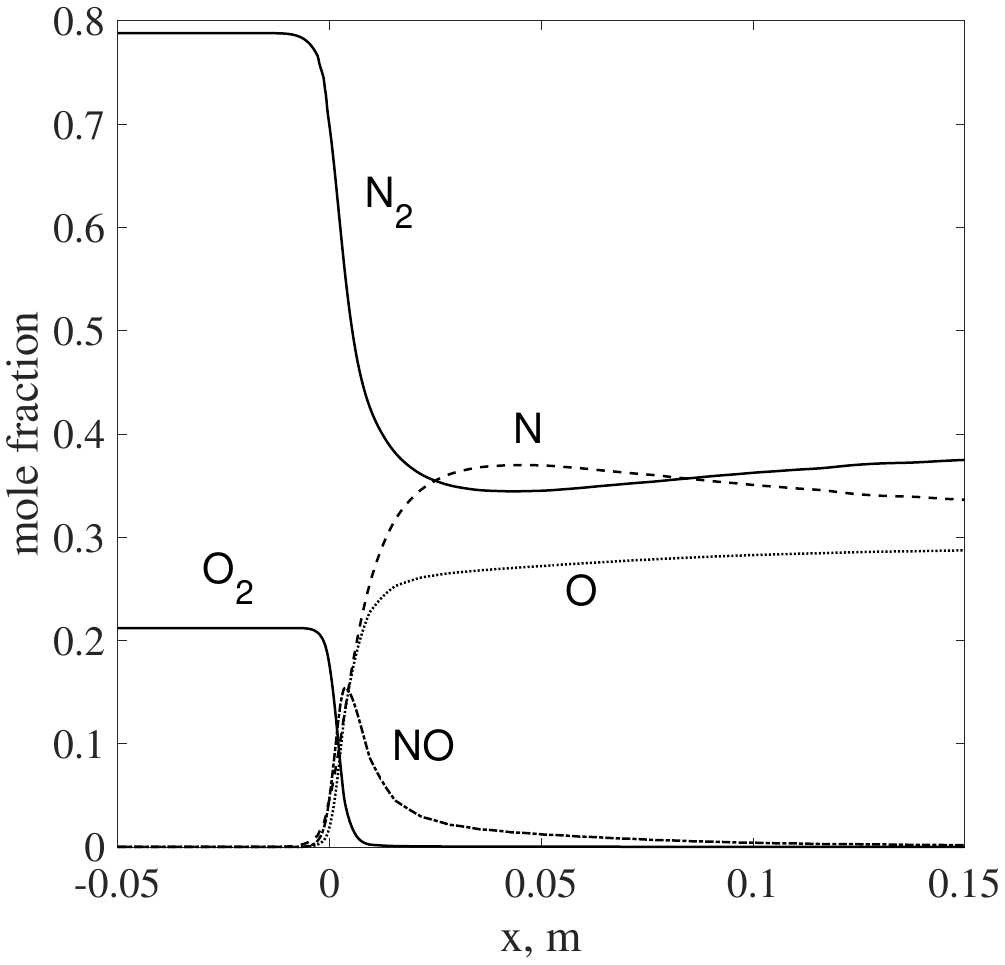}
     \figurecaption{Neutral species mole fractions along a representative streamline located 6~cm above the stagnation streamline for the 61~km altitude case.}
     \label{fig:species_streamline3}
\end{figure}

\begin{figure}[!h]
     \centering
     \subfigure[61 km]{~~~\includegraphics[width=0.36\textwidth]{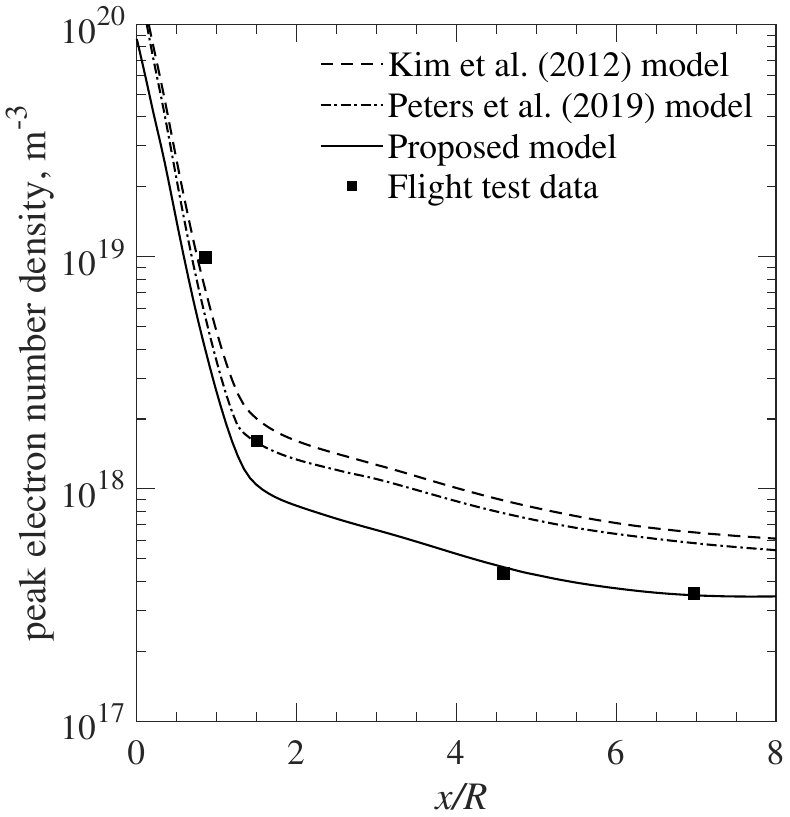}~~~}
     \subfigure[71 km]{~~~\includegraphics[width=0.36\textwidth]{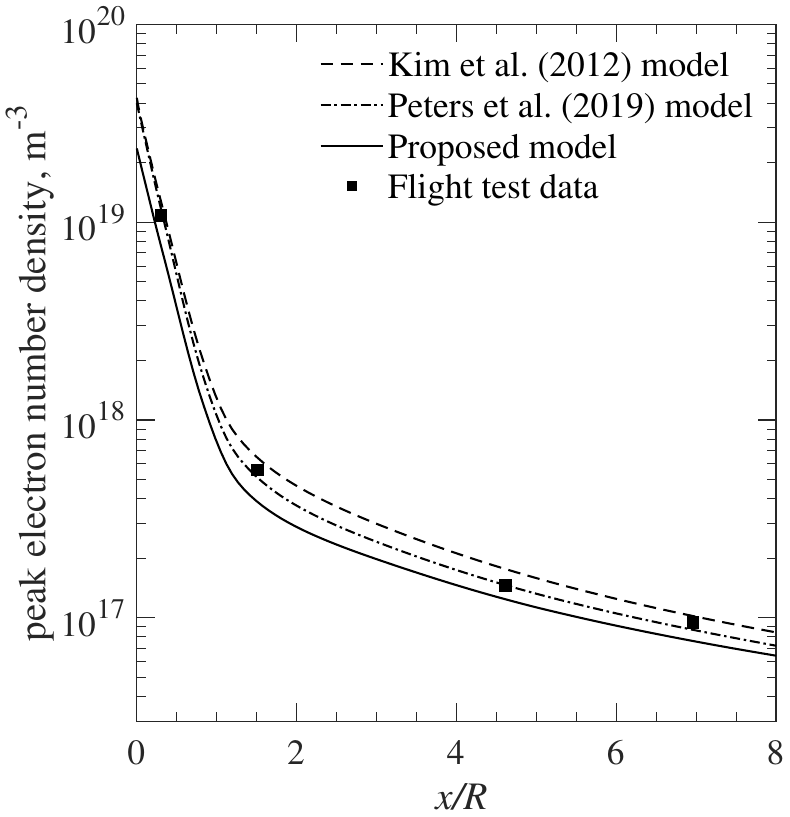}~~~}
     \figurecaption{Comparison between different inelastic heating models and RAM-C-II flight test data on the basis of peak electron number density along the
vehicle axis at (a) 61 km altitude and (b) 71 km altitude.}
     \label{fig:RAMCII_Nemax_comparison}
\end{figure}

\subsection{RAM-C-II Re-entry Flight Test Case}

As outlined in \cite{nasa:1970:akey}, the RAM-C-II flight test was conducted in the late 60s to investigate ionization around the plasma flow surrounding a reentry vehicle at a freestream velocity of around 7.65 km/s. The vehicle is a 1.2-meter long body with a blunt leading-edge radius of 15.24 cm and a 9 deg cone half-angle, shown in Fig.~\ref{fig:setup_RAMCII}.  In the second flight test electron density was measured at altitudes of 61 and 71 kilometers where the freestream static pressure corresponds to 19.62 and 4.844 Pa, respectively.

An assessment of grid-induced numerical error is shown in Fig.\ \ref{fig:RAMCII_grid_convergence_Ne} for the 61 km altitude reentry case. The mesh is refined by doubling the number of grid lines in each dimension starting from a coarser $125\times90$ mesh. Using the finest $500\times360$ grid is seen to result in a numerical error of less than $5$\% on the maximum electron number density along the stagnation line.

In Fig.~\ref{fig:RAMCII_Tv_comparison}, a comparison of vibrational temperature around the RAM-C-II body is shown for altitudes of 61 and 71 kilometers. Corresponding electron temperature contours are shown in Fig.~\ref{fig:RAMCII_Te_comparison} obtained with different inelastic electron heating models. In the case of the \cite{jtht:2012:kim} model, the electron temperature quickly relaxes towards the vibrational temperature  (see Fig.~\ref{fig:RAMCII_Tv_comparison} and Fig.~\ref{fig:RAMCII_Te_comparison}a,b); in contrast, the \cite{jap:2019:peters} model and the proposed inelastic electron heating model exhibit a large difference between vibrational and electron temperature by a factor of 5 to 6. The reason why the \cite{jtht:2012:kim} model leads to very fast relaxation of electron temperature towards the vibrational temperature at both altitudes is that the latter model has (i) a higher heating over cooling ratio than the other models as discussed previously in Section III.D and (ii) stronger relaxation terms compared to other models as discussed in \cite{pf:2024:parent}.

A key finding is that the proposed model predicts a significantly lower electron temperature than the \cite{jap:2019:peters} model, as shown in Fig.~\ref{fig:RAMCII_Te_comparison}. Although both models are derived from the detailed balance principle, this discrepancy in temperature stems from a crucial difference in how the total electron energy loss is calculated. The primary cause is our model's inclusion of inelastic cooling from all neutral species, whereas the \cite{jap:2019:peters}  model accounts only for cooling from N$_2$. In the high-temperature region behind the bow shock, a significant amount of nitric oxide (NO) is formed. This NO provides an additional pathway for electron cooling—an effect not present in the \cite{jap:2019:peters} model and which leads to a lower overall electron temperature.

{

A critical modeling choice is whether to include vibrational heating from nitric oxide (NO). While the classical \cite{jcp:1963:millikan} correlation suggests a large relaxation distance that would justify its exclusion, we acknowledge that recent experimental work indicates much faster relaxation; this discrepancy highlights a key uncertainty in current models. To assess this, we first note the conditions from our simulation:} substantial amounts of NO exist only within a narrow 1-2 cm region behind the bow shock (Fig.~\ref{fig:species_streamline3}), where the flow velocity is approximately 2700~m/s. Applying the classical Millikan-White model to these conditions (using a representative gas temperature of 13,000~K and an NO partial pressure of 800~Pa) predicts a vibrational relaxation distance of about 5~cm. This distance—governed by the rapid NO--NO V-T relaxation process rather than the much slower N$_2$--NO V-T process—is several times larger than the zone of NO abundance. { The classical analysis therefore implies the molecules pass through the region too quickly to become vibrationally hot, acting primarily as an electron cooling sink.

However, this conclusion based on the classical model is challenged by recent findings. Experiments by \cite{pof:2022:streicher} indicate that the relaxation distance predicted by the Millikan-White correlation is overestimated and could be as little as 7 mm. Moreover, recent simulations by \cite{aiaapaper:2018:andrienko} and \cite{jtht:2025:thirani} show that NO--O relaxation is a faster process than the NO--NO relaxation considered here, potentially leading to even smaller relaxation distances. If these faster rates hold, the vibrational temperature of NO may approach the translational temperature of the heavy species. Consequently, substantial electron heating from vibrationally excited NO would occur, leading to a higher electron temperature. For this work, we proceed by excluding NO heating to isolate the effects of the N$_2$ model, but recognize that incorporating a multi-vibrational temperature model for NO represents a critical path for future research.}

The effect of these electron temperature differences on the plasma density can be assessed by measuring the maximum electron density along the vehicle axis as shown Fig.~\ref{fig:RAMCII_Nemax_comparison}. Here the axial distance $x$ is normalized with the leading edge radius $R$ such that $x/R = 0$ corresponds to the stagnation point. We obtain better agreement to flight test data at 61 km using the proposed electron heating model with an error being within a few percents far from the leading edge. Much larger discrepancy to experimental data is found using the \cite{jtht:2012:kim} and \cite{jap:2019:peters} models, with the latter models overpredicting peak plasma density by 80\% and 98\% respectively further downstream of the RAM-C-II leading edge. Such differences are mostly due to the decrease in electron-ion recombination rates as a result of the higher electron temperature.

\begin{figure}[!b]
     \centering
     \subfigure[]{\includegraphics[width=0.13\textwidth]{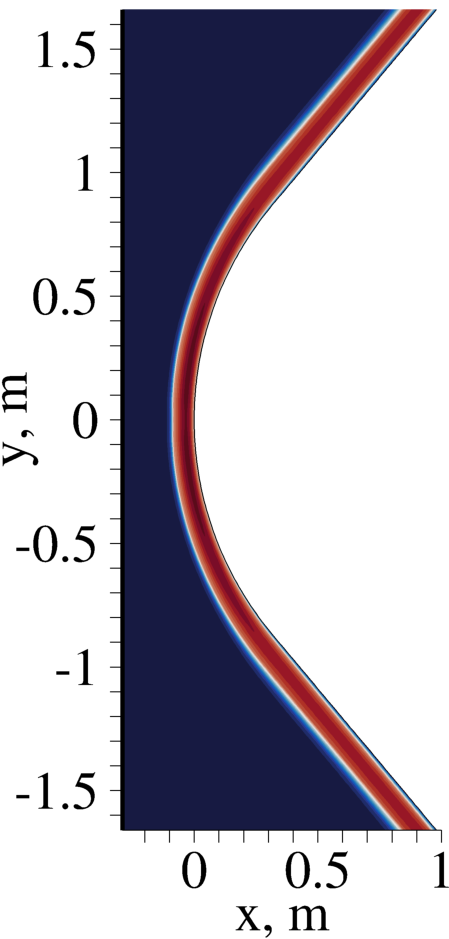}}
     \subfigure[]{\includegraphics[width=0.13\textwidth]{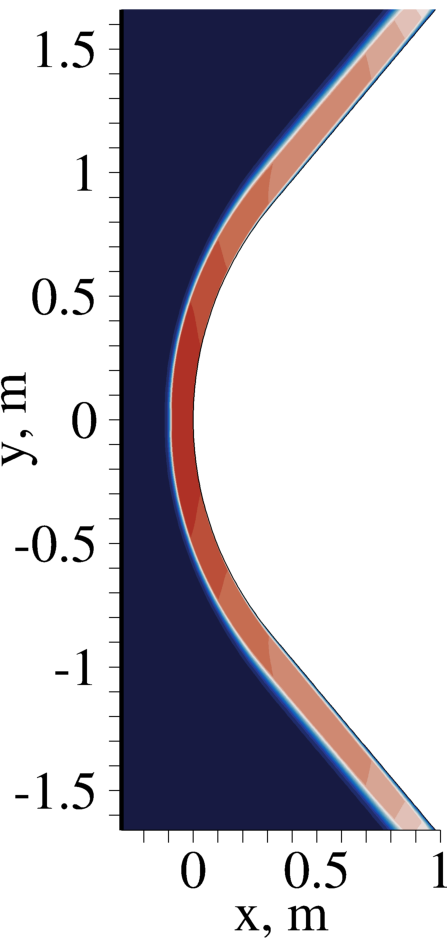}}
    \subfigure[]{\includegraphics[width=0.188\textwidth]{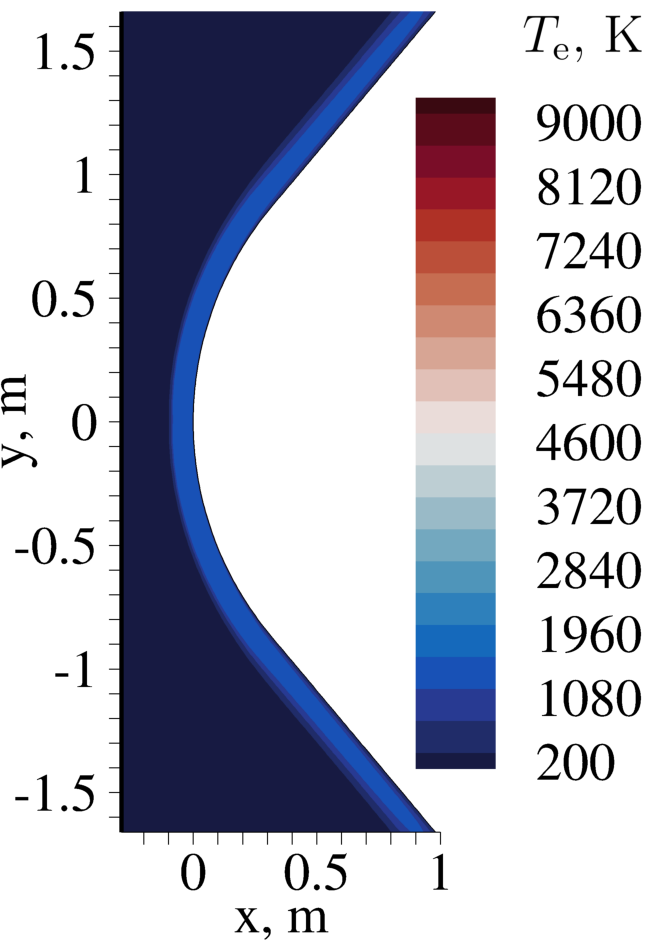}}
     \figurecaption{Electron temperature around the OREX body for the 84 km altitude case obtained with different electron heating
models: (a) Kim et al. model, (b) Peters et al. model and (c) proposed model.}
     \label{fig:Te_contours_OREX}
\end{figure}

\begin{figure}[!b]
     \centering
     \includegraphics[width=0.38\textwidth]{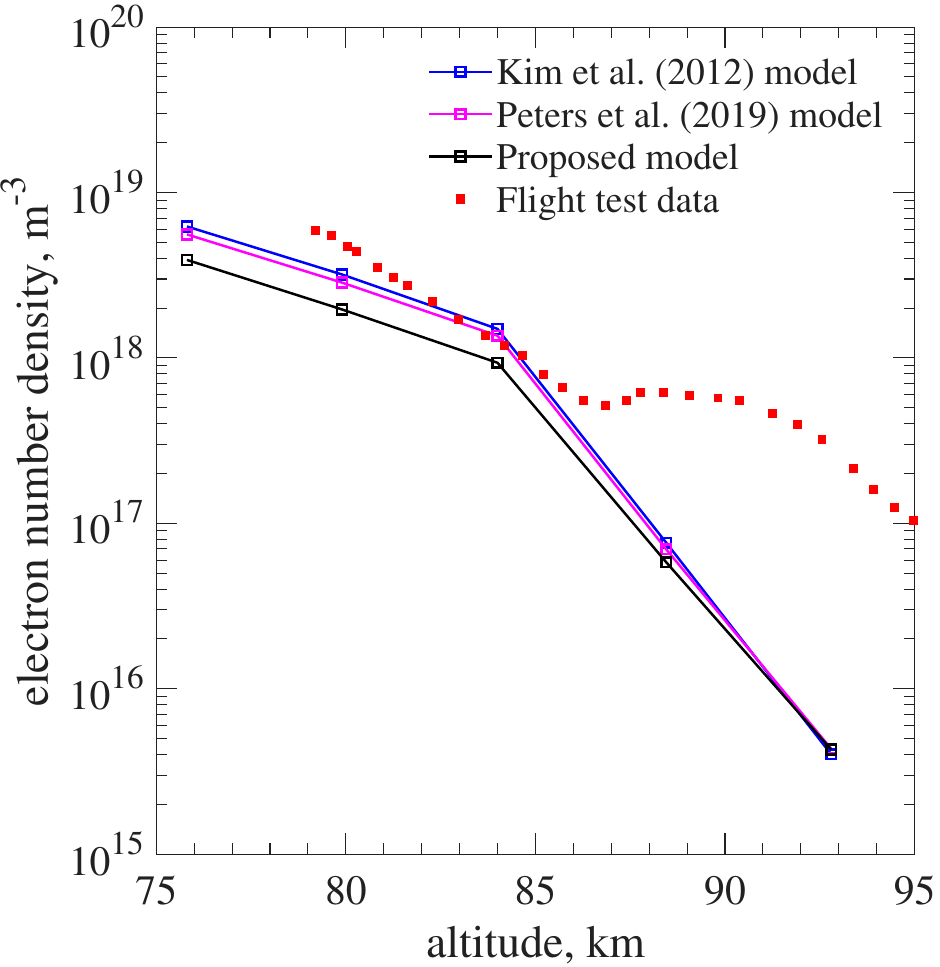}
     \figurecaption{Comparison between different inelastic heating models and
OREX flight test data on the basis of electron number density measured at probe number 3.}
     \label{fig:OREX_Ne_probe3}
\end{figure}

\begin{figure}[!b]
     \centering
     \includegraphics[width=0.38\textwidth]{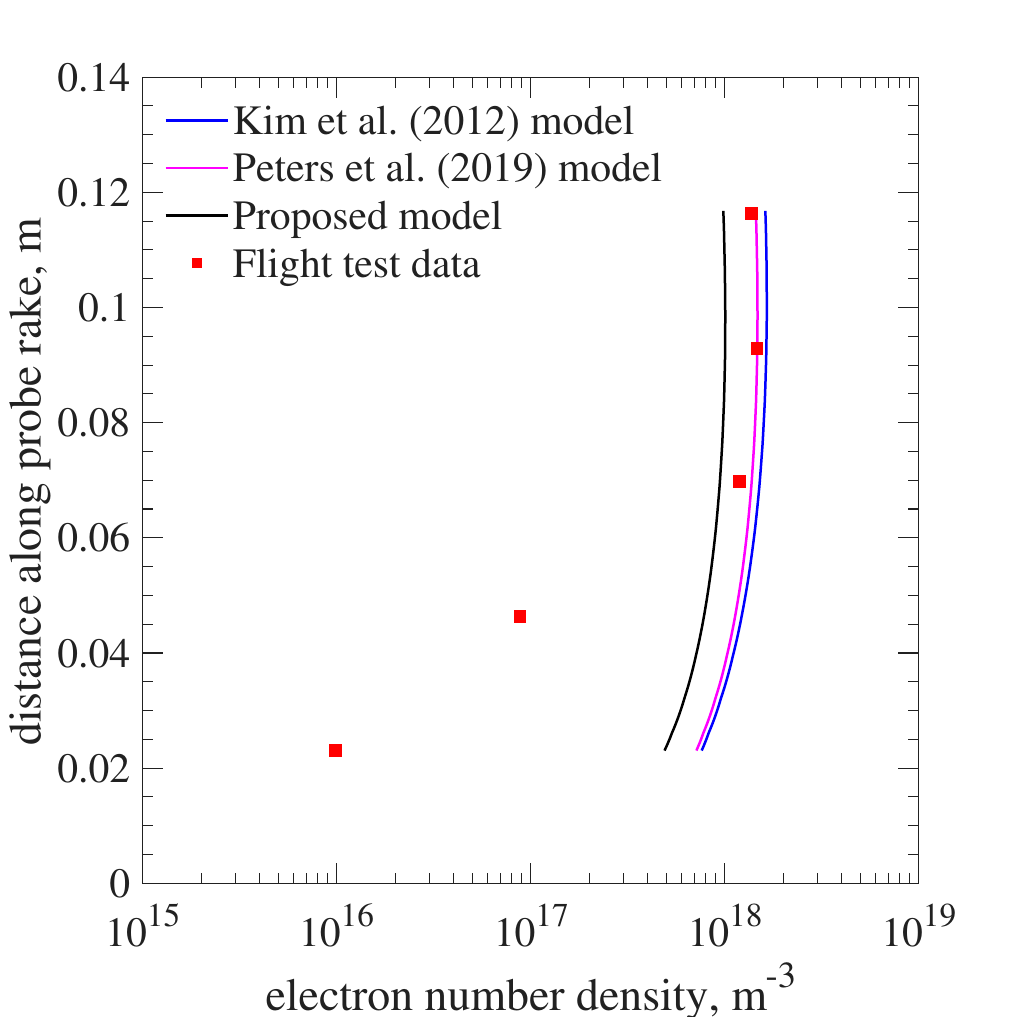}
     \figurecaption{Comparison between different inelastic heating models and
OREX flight test data on the basis of electron number density measured along the probe rake at an altitude of 84 km.}
     \label{fig:OREX_Ne_84km}
\end{figure}

\begin{figure}[!t]
     \centering
     \includegraphics[width=0.38\textwidth]{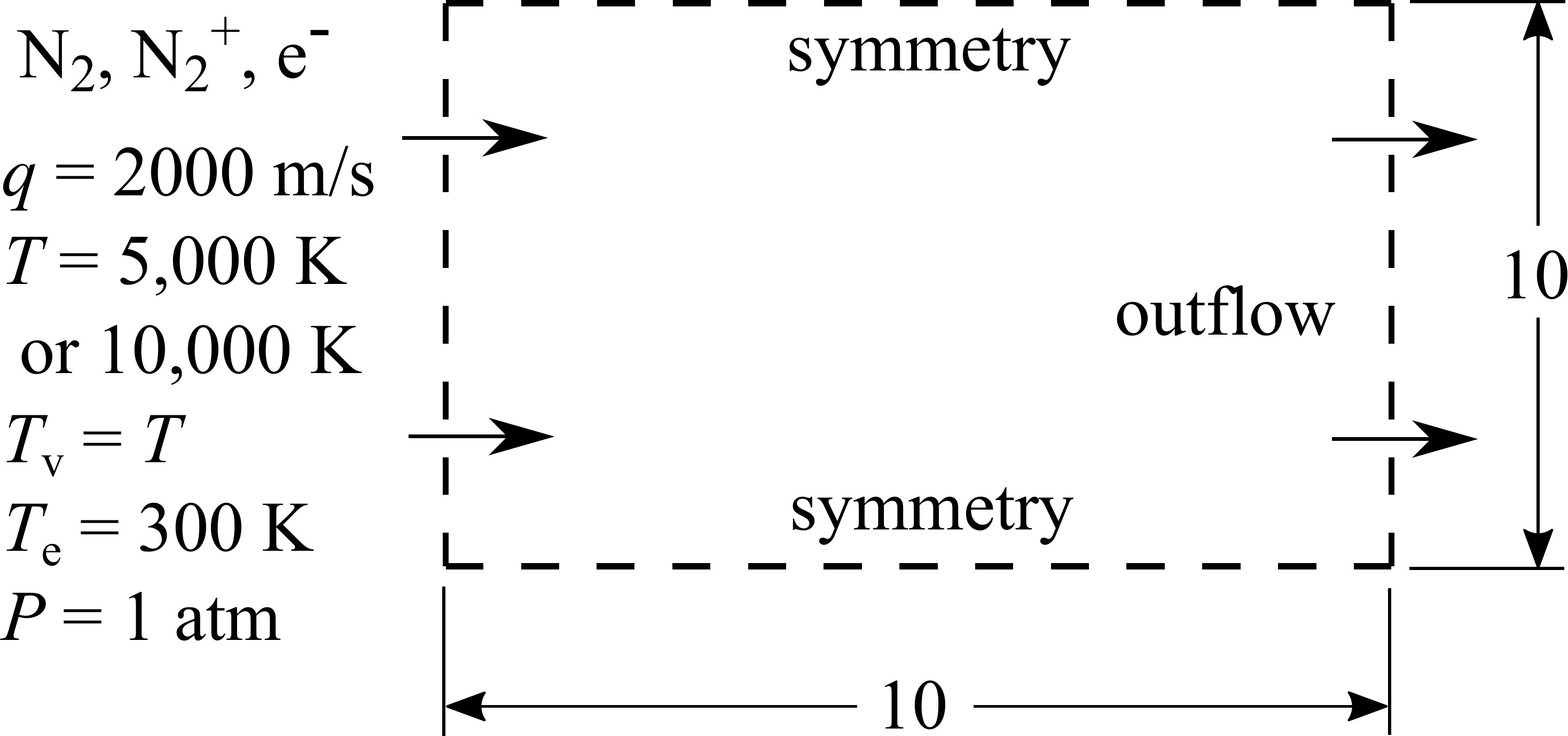}
     \figurecaption{Problem setup for the thermal equilibrium test case; dimensions in meters.}
     \label{fig:setup_equil}
\end{figure}

At 71 km altitude, we again observe significant changes in the electron temperature between models similar to the 61~km case. However, the electron density here is not as sensitive to electron temperature, and much better agreement to flight test data can be observed across all models. In this case all models perform similarly and lead to a maximum 20\% discrepancy to flight test measurements.

The small discrepancy between the proposed model and the flight test data could be due errors in reflectometer measurements. Indeed, experimental measurements of electron density inferred from the microwave reflectometers report a factor-of-2 error bar as discussed in \cite{nasa:1970:grantham}, partly attributed to phase errors due to probe motion. 
~\\

\subsection{OREX Re-entry Flight Test Case}

\cite{asvpaper:1995:inouye} details the Orbital Reentry Experiment (OREX), which was conducted in 1994 by the National Aerospace Laboratory and the National Space Development Agency of Japan. The vehicle's geometry consists of a 1.35~m radius nose followed by a 50-degree half-cone wedge. An electrostatic probe rake was mounted on the vehicle, 1.53~meters behind the leading-edge stagnation point. Along this rake, five probes measured saturated ion currents. These measurements were used to infer the electron number density during flight at altitudes between 80 and 100 kilometers.

As with the previous reentry test case, the proposed electron heating model leads to a significantly lower electron temperature around the OREX vehicle. This result is compared to the models from \cite{jap:2019:peters} and \cite{jtht:2012:kim}, with the electron temperature contours shown in Fig.~\ref{fig:Te_contours_OREX}. The proposed model includes electron cooling from NO, which is still present in significant amounts behind the bow shock. This inclusion leads to an electron temperature up to six times lower than that predicted by the \cite{jap:2019:peters} model, which only accounts for inelastic energy losses to N$_2$ and the corresponding energy gains.

In Fig.~\ref{fig:OREX_Ne_probe3}, we evaluate this modeling choice against OREX flight data using electron number density measurements from probe 3 at several altitudes. The results show good qualitative agreement with flight data for altitudes below 84 km. The proposed model yields a reduction in electron number density, which is likely a result of higher electron-ion recombination rates caused by the lower electron temperature. Above 84 km, the sharp decline in electron density is attributed to the Park-like chemical model's use of $\sqrt{T T_{\rm v}}$ as the controlling temperature for several ionization reactions, rather than just $T$. As discussed in \cite{pf:2022:parent}, this choice can be inaccurate at such high altitudes and lead to too little ionization because it reduces too strongly the rate of associative ionization of NO.
\begin{table*}[!ht]
\center\fontsizetable
  \begin{threeparttable}
\renewcommand{\arraystretch}{1.3}
\tablecaption{Effect of electron temperature models on measured vibrational and electron temperatures at thermal equilibrium.}
\fontsizetable
\begin{tabular*}{\textwidth}{@{}l@{\extracolsep{\fill}}cccc@{}}
\toprule
~  & \multicolumn{2}{c}{Inflow $T=5000$ K}  & \multicolumn{2}{c}{Inflow $T=10000$ K}\\
\cmidrule(lr){2-3}\cmidrule(lr){4-5}
Electron temperature model  & Equilibrium $T_{\rm v}$, K   &Equilibrium $T_{\rm e}$, K  & Equilibrium $T_{\rm v}$, K    & Equilibrium $T_{\rm e}$, K  \\
\midrule
\cite{jap:2010:shneider}, Eq.~(\ref{eqn:scale_TediffTv}) & $5000$  & $5000$  & $10000$ & $10000$    \\
 \cite{jtht:2012:kim}, Eq.~(\ref{eqn:scale_LandauTeller}) & $5000$  & $5000$  & $10000$ & $10000$   \\
\cite{jap:2019:peters}, Eq.~(\ref{eqn:scale_Peters}) & $4990$  & $3356$  & $9990$ & $4770$     \\
Proposed model, Eq.~(\ref{eqn:scale_proposed})  & $5000$  & $5000$  & $10000$ & $10000$    \\

\bottomrule
\end{tabular*}
\label{tab:equilbriumcheck}
\end{threeparttable}
\end{table*}

For the 84 km altitude case, Fig.~\ref{fig:OREX_Ne_84km} shows another comparison to flight data, this time covering electron density data from all five probes along the rake. Good agreement is obtained with measurements from probes 3 to 5, which are located further from the OREX surface. However, the simulation results do not capture the data from the near-wall probes (1 and 2). This discrepancy may be partly due to errors in post-processing the electron density data from the original ion current measurements.

\subsection{Thermal Equilibrium Test Case}

The fourth test case is an assessment of thermal equilibrium comparing the different inelastic electron heating models outlined previously. A pre-ionized flow enters the computational domain shown in Fig.\ \ref{fig:setup_equil} with a speed of 2000 m/s, a temperature of 5,000 K or 10,000 K and a pressure of 1 atm. The electron molar fraction is of 0.01, the molar fraction of $\rm N_2^+$ is 0.01 and the $\rm N_2$ molar fraction is 0.98. The 2D computational domain is 10 meters in length and 10 meters in height. Chemical reactions are disabled to ensure constant densities in the domain.

From the converged steady-state solutions we measure the gas temperature, the vibrational temperature and the electron temperature near the supersonic outflow boundary. These measurements are presented in Table\ \ref{tab:equilbriumcheck} comparing the results of three previous electron heating models to the model proposed herein.

The phenomenological model {(one based on observed phenomena rather than derived from first principles)} by \cite{jap:2010:shneider} where the net heating-cooling rate scales as $T_{\rm e}-T_{\rm v}$  and the \cite{jtht:2012:kim} model both quickly equilibrate electron and vibrational temperatures. In contrast, the \cite{jap:2019:peters} does not ensure $T_{\rm e}=T_{\rm v}$ at equilibrium. Although this model satisfies the detailed balance principle, it does not guarantee the equality of electron and vibrational temperatures at equilibrium (see Fig.~\ref{fig:contours_lossesgains}c), even more so as the gas inflow temperature is increased. The model proposed herein performs as expected by giving $T_{\rm e}=T_{\rm v}$ with a very small difference to the set gas inflow temperatures.

\section{Conclusions}

A new, thermodynamically consistent model for the vibrational-electron heating term  has been developed and validated. This novel approach to modeling e-V relaxation satisfies the principle of detailed balance and, crucially, ensures the correct equilibration of electron and vibrational temperatures in plasmas approaching thermal equilibrium.

A key advantage of this model is its formulation, which obtains {V-e heating by multiplying the e-V cooling rate} by a simple algebraic factor dependent only on the electron and vibrational temperatures. This allows the model to be coupled with any cooling description, including those derived from swarm experiments. Using swarm data is particularly advantageous because it correctly accounts for all inelastic energy losses as a function of electron temperature. This avoids the need for separate and often incomplete cross-sectional data required by other methods.

The proposed model is validated against several test cases, including the non-equilibrium plasma flow around the RAM-C-II flight vehicle. In that case, the model predicts an electron temperature several times lower than previous models, which increases electron-ion recombination rates and resulted in improved agreement with flight test data. The model is also shown to be thermodynamically consistent in a dedicated thermal equilibrium test case.

{

A central assumption of this model is that the energy exchange in electron cooling corresponds to the fundamental vibrational transition. This approximation holds for electron temperatures below 1~eV characteristic of hypersonic flows, where excitations leading to overtones contribute minimally to the total cooling. However, in applications like plasma-assisted combustion or laser-induced plasmas where electron temperatures can reach 3–6~eV, this assumption fails. For these scenarios, a hybrid approach is recommended: employing a more detailed overtone model at high temperatures and switching to the present thermodynamically consistent model below 1~eV. This strategy is critical for accurately simulating the cool-down phase between high-energy pulses, which governs plasma recombination and the conditions for subsequent breakdown.

Another potential source of error is the model's assumption of a Boltzmann distribution for vibrational energy. However, this approach is broadly justified because rapid V-V relaxation quickly restores a Boltzmann shape under most conditions. An alternative state-to-state model, while physically more detailed, is often impractical due to uncertain rate data and a prohibitive computational cost that can introduce larger numerical errors. Thus, assuming a Boltzmann distribution frequently represents a superior trade-off, minimizing the total simulation error.

Improved accuracy and consistency in this electron energy model have significant implications across the diverse fields where e-V coupling is critical. In hypersonic and re-entry flows, it enables more reliable predictions of plasma conductivity and density, which are fundamental to developing technologies like MHD systems, electron transpiration cooling, and electromagnetic shielding. For plasma-assisted combustion, the model provides a more predictive foundation for simulating slow e-V energy exchange pathways, moving beyond approximate energy partitioning factors. In the study of laser-induced plasmas, it allows for more accurate modeling of the plasma's temporal evolution and lifetime, which is governed by electron temperature-dependent recombination and attachment rates. Ultimately, by providing a robust and physically-sound closure, this work can accelerate the design and analysis of advanced systems across all these plasma-based applications.}

\footnotesize
\bibliography{all}
\bibliographystyle{plainnatmod}
\end{document}